\documentclass[reqno,10pt,a4paper,table]{amsart}

\numberwithin{equation}{section}
\allowdisplaybreaks 

\usepackage{mathtools} 
\usepackage{amssymb,eucal,mathrsfs,setspace}
\usepackage[noadjust]{cite}
\usepackage{graphicx}
\usepackage{array}
\newcolumntype{C}{>{$}c<{$}} 

\usepackage[largesc,theoremfont]{newtxtext}       
\usepackage[libertine,cmbraces,varbb]{newtxmath}  

\begingroup
\makeatletter
\@for\theoremstyle:=definition,remark,plain\do{%
\expandafter\g@addto@macro\csname th@\theoremstyle\endcsname{%
\addtolength\thm@preskip{.5\baselineskip plus .2\baselineskip minus .2\baselineskip}
\addtolength\thm@postskip{.5\baselineskip plus .2\baselineskip minus .2\baselineskip}
}%
}
\endgroup

\newcommand{\wun}{\vvmathbb{1}}  

\DeclareSymbolFont{largesymbolspal}{OMX}{zplm}{m}{n} 
\let\oint\relax
\DeclareMathSymbol{\ointop}{\mathop}{largesymbolspal}{"49}
\newcommand{\oint}{\ointop\nolimits}

\usepackage[inner=25mm, outer=25mm, top=25mm, bottom=25mm, head=10mm, foot=10mm]{geometry}

\usepackage{enumitem}
\setlist[itemize]{leftmargin=*}               
\setlist[enumerate]{leftmargin=*,             
	label=\textup{(\roman*)}}                   

\usepackage[colorlinks=true,citecolor=red,linkcolor=blue]{hyperref} 

\usepackage[capitalise,noabbrev]{cleveref} 

\usepackage{tikz}

\newcommand{\pd}{\partial}              

\newcommand{\vacuum}{\Omega}  
\newcommand{\confvec}{\omega} 

\renewcommand{\ge}{\geqslant} 
\renewcommand{\le}{\leqslant} 

\newcommand{\dd}{\mathrm{d}}   
\newcommand{\ee}{\mathsf{e}}   
\newcommand{\ii}{\mathfrak{i}}   

\DeclarePairedDelimiter{\brac}{\lparen}{\rparen}   
\DeclarePairedDelimiter{\sqbrac}{\lbrack}{\rbrack} 
\DeclarePairedDelimiter{\set}{\lbrace}{\rbrace}
\newcommand{\st}{\mspace{5mu} {:} \mspace{5mu}}    
\DeclarePairedDelimiter{\abs}{\lvert}{\rvert}

\DeclarePairedDelimiter{\ang}{\langle}{\rangle}    

\DeclarePairedDelimiter{\normord}{{:}}{{:}}        

\DeclarePairedDelimiterX{\comm}[2]{\lbrack}{\rbrack}{#1 , #2}  
\DeclarePairedDelimiterX{\acomm}[2]{\lbrace}{\rbrace}{#1 , #2} 
\DeclarePairedDelimiterX{\inner}[2]{\langle}{\rangle}{#1 , #2} 
\DeclarePairedDelimiterX{\super}[2]{\lparen}{\rparen}{#1 \delimsize\vert \mathopen{} #2} 

\renewcommand{\implies}{\Rightarrow}

\newcommand{\lra}{\longrightarrow}
\newcommand{\ira}{\hookrightarrow}    
\newcommand{\lsra}{\ensuremath{\relbar\joinrel\twoheadrightarrow}}        

\newcommand{\dses}[3]{0 \lra #1 \lra #2 \lra #3 \lra 0} 

\newcommand{\categ}[1]{\mathscr{#1}} 

\newcommand{\fld}[1]{\mathbb{#1}}    
\newcommand{\ZZ}{\fld{Z}}

\newcommand{\QQ}{\fld{Q}}
\newcommand{\RR}{\fld{R}}
\newcommand{\CC}{\fld{C}}

\newcommand{\grp}[1]{\mathsf{#1}}              
\newcommand{\SLG}[2]{\grp{#1}(#2)}             

\newcommand{\alg}[1]{\mathfrak{#1}}                      
\newcommand{\finite}[1]{#1}                              

\newcommand{\SLA}[2]{\finite{\alg{#1}}{}_{#2}}           

\newcommand{\ideal}[1]{\ang*{#1}}

\newcommand{\cox}{\mathsf{h}}                   
\newcommand{\dcox}{\cox^{\vee}}                 

\newcommand{\fuse}{\mathbin{\times}}                              
\newcommand{\fuscoeff}[3]{\mathcal{N}_{#1#2}^{\hphantom{#1#2}#3}} 

\newcommand{\coprosymb}{\Delta}
\newcommand{\coprow}[1]{\coprosymb_{#1}}			                
\newcommand{\coproi}[2]{\coprosymb^{(#1)} \brac[\big]{#2}}         
\newcommand{\copropts}[2]{\coprosymb_{#1} \brac[\big]{#2}}         
\newcommand{\coproipts}[3]{\coprosymb^{(#1)}_{#2} \brac[\big]{#3}} 
\newcommand{\coprono}[2]{\coprosymb^{(#1)}_{#2}}             

\newcommand{\cft}{conformal field theory}
\newcommand{\cfts}{conformal field theories}

\newcommand{\voa}{vertex operator algebra}
\newcommand{\voas}{vertex operator algebras}
\newcommand{\vosa}{vertex operator subalgebra}

\newcommand{\ope}{operator product expansion}
\newcommand{\opes}{operator product expansions}

\newcommand{\wzw}{Wess-Zumino-Witten}
\newcommand{\ngk}{Nahm-Gaberdiel-Kausch}

\newcommand{\hw}{highest-weight}
\newcommand{\hwv}{\hw{} vector}
\newcommand{\hwvs}{\hw{} vectors}
\newcommand{\hwm}{\hw{} module}
\newcommand{\hwms}{\hw{} modules}
\newcommand{\sv}{singular vector}
\newcommand{\svs}{singular vectors}
\newcommand{\lhs}{left-hand side}

\newcommand{\rhs}{right-hand side}

\newcommand{\fus}[1]{\mathbin{\boxtimes_{#1}}}
\newcommand{\opp}{{\mathsf{opp}}}
\newcommand{\topspace}{{\mathsf{top}}}

\newcommand{\intwtype}[3]{\begin{psmallmatrix} #1 \\ #2 \, #3 \end{psmallmatrix}} 

\newcommand{\Res}[3]{\oint_{#2} #3 \, \frac{\dd #1}{2 \pi \ii}}

\theoremstyle{definition}
\newtheorem{rema}{Remark}[section]
\newtheorem{dfn}[rema]{Definition}
\newtheorem{exam}{Example}
\newtheorem{example}[exam]{Example}

\theoremstyle{plain}

\title{NGK and HLZ: fusion for physicists and mathematicians}

\author[S~Kanade]{Shashank Kanade}
\address[Shashank Kanade]{
Department of Mathematics \\
University of Denver \\
Denver, USA, 80208.
}
\email{shashank.kanade@du.edu}

\author[D~Ridout]{David Ridout}
\address[David Ridout]{
School of Mathematics and Statistics \\
University of Melbourne \\
Parkville, Australia, 3010.
}
\email{david.ridout@unimelb.edu.au}

\allowdisplaybreaks

\begin{document}

\maketitle

\begin{abstract}
	In this expository note, we compare the fusion product of conformal field theory, as defined by Gaberdiel and used in the Nahm-Gaberdiel-Kausch (NGK) algorithm, with the $P(w)$-tensor product of vertex operator algebra modules, as defined by Huang, Lepowsky and Zhang (HLZ). We explain how the equality of the two ``coproducts'' derived by NGK is essentially dual to the $P(w)$-compatibility condition of HLZ and how the algorithm of NGK for computing fusion products may be adapted to the setting of HLZ.
	We provide explicit calculations and instructive examples to illustrate both approaches.
	This document does not provide precise descriptions of all statements, it is intended more as a gentle starting point for the appreciation of the depth of the theory on both sides.
\end{abstract}

\onehalfspacing

\bigskip

\setcounter{tocdepth}{1}
\tableofcontents

\section{Before we begin}

\bigskip

\begin{quote}
\textsl{``Perhaps fusion always has the quasi-rational features seen for the free bosons. It will be important to investigate Calabi-Yau spaces from this point of view, but the tools for this study have yet to be developed. As a first step, we need a rigorous and convenient definition of the fusion product for generic theories, and better algorithms for its evaluation.''
}
\quad\dots{Werner Nahm \cite{NahQua94}}
\end{quote}

\bigskip

\subsection{Why?}

It is a highly non-trivial matter to form tensor products (called fusion products in physics parlance) of modules for a given vertex operator algebra (chiral algebra), not to mention building a braided tensor category out of these modules.

The late 80s and early 90s witnessed an intense period of activity devoted to this problem:
\begin{itemize}
	\item Feigin and Fuchs described the fusion coefficients of certain \cfts{} as dimensions of spaces of coinvariants \cite{FeiCoh88}.
	\item Moore and Seiberg wrote their highly influential papers \cite{MooSeiPoly88,MooSeiCFT89} on rational conformal field theories,\footnote{A \cft{} is said to be \emph{rational} if its quantum state space is semisimple (completely reducible), decomposing into a finite direct sum of tensor products of irreducible modules.  The name reflects the fact that such theories have rational central charges and conformal weights.} introducing a ``coproduct-like'' formula for the action of the chiral algebra.  However, they incorrectly identified the vector space underlying the fusion product as that underlying the usual tensor product.
	\item Frenkel and Zhu codified the coinvariant approach for certain classes of modules over quite general \voas{} \cite{FreVer92}, see also work of Li \cite{LiThesis94,LiNuc98}, obtaining a formula for the fusion coefficients in the rational case.
	\item Gaberdiel extended the coproduct formula of Moore-Seiberg and corrected their work by using locality to (morally) define the fusion product as a quotient of the vector space tensor product \cite{GabFus94,GabFus94b}.
	\item Kazhdan and Lusztig proved \cite{KazAff91,KL1,KL2,KL3,KL4} major theorems defining fusion rigorously and relating certain tensor categories for affine Lie algebras at non-rational levels to quantum group tensor categories.
	\item Nahm introduced an algorithmic method to analyse and, in favourable cases (rational \cfts{} in particular), identify fusion products \cite{NahQua94}.
	\item Huang and Lepowsky wrote their first series of papers \cite{HL-vtc,HuaTheI95,HuaTheII95,HuaTheIII95} on rigorously defining fusion and proving tensor structure theorems for appropriate module categories over rational vertex operator algebras.
	\item Gaberdiel and Kausch extended Nahm's methods to include logarithmic \cfts{}\footnote{A \cft{} is said to be \emph{logarithmic} if its quantum state space is not completely reducible.  The name reflects the fact that all known examples possess a non-diagonalisable action of the hamiltonian which leads to logarithmic singularities in certain correlation functions.} and implemented them on a computer \cite{GabInd96}.  The resulting algorithm is now known as the \ngk{} fusion algorithm.
\end{itemize}
We will not discuss this history in any more detail, instead referring the reader to resources such as \cite{HuaLepSurvey13} (for mathematicians) and \cite{GabInt00,CreLog13} (for physicists), as well as to the references cited therein.

Our aim in this expository note is to focus exclusively on the physicists' computational approach, as explained by Nahm, Gaberdiel and Kausch (NGK henceforth), and the mathematicians' rigorous approach, as developed by Huang, Lepowsky and Zhang (HLZ for short) in \cite{HLZ}. Here, we concentrate only on the definition of the fusion product and its algorithmic construction; the question of whether it is possible to build a braided tensor category from this fusion product is much much more difficult.  Readers interested in the categorical structures underlying \cft{} can instead turn to papers such as \cite{KirqAn02,HuaBra15,CreTen17}.

The approaches of NGK and HLZ start from quite different points of view and use very different language.  However, the ingredients are almost identical.  The main idea is to somehow construct the fusion product out of the vector space tensor product of modules.  Along the way, one has to deal with some nasty convergence issues and to make everything work smoothly, a fix is needed.  Physicists were well aware of this difficulty, see \cite{GabFus94,NahQua94}, and the NGK fusion algorithm avoids these convergence issues by working with various quotients of the fusion product.  On the other hand, Huang and Lepowsky successfully tackled this issue rigorously for rational models by passing to the dual space and, together with Zhang, subsequently extended their dual space formalism to cover logarithmic cases.

In this note, we wish to explain two main points. First, the NGK ``definition'' of the fusion product rests on imposing the condition that two seemingly different coproducts are in fact the same, while the HLZ tensor product is built from functionals in the dual space that are required to satisfy a certain compatibility condition. We will show that these two conditions are essentially the same thing (more precisely, they are duals of one another).
Second, we will focus on algorithmic implementations for constructing fusion products. We shall explain this with the aid of a specific example in both the NGK and HLZ formalisms.

Our purpose is to facilitate a dialogue between mathematicians and physicists by presenting a coherent fusion (pun intended) of various ideas.  In particular, we would like to reassure mathematicians that the fusion rules that physicists compute with NGK can be rigorously justified (in principle) and also to reassure physicists that the theorems that mathematicians prove with the universal tensor product theory of HLZ are indeed results about fusion.  We would however like to remind the reader that this paper should not be relied upon for precise statements of results.  While we have done our best not to tell any outright lies, the theory of fusion is notoriously subtle and a full account with all details explained would necessarily take more than the space we have here.  As always, the cited literature is the canonical source for these details.

\subsection{How?}

This note is organised as follows.

In \cref{sec:language}, we first provide a review of the (sometimes wildly different) notation and terminology used by mathematicians and physicists, so as to make it easier for both audiences to follow the rest of the paper.  We also take the opportunity to fix some of our own choices in this regard.  Our compromises will no doubt outrage many readers, but we take solace in the fact that it is really not possible to please everyone in this respect.

We begin our exposition by first presenting Gaberdiel's original approach to defining fusion products in \cref{sec:NGK}. Here, we will explain how the fusion product of two modules is morally constructed from their vector space tensor product by imposing a certain equality of two coproducts.  These coproducts are derived from locality, so the definition appears natural and uncontroversial.  However, we point out concretely that this equality of coproducts does not make sense in general because it involves infinite coefficients when expanded in the usual fashion.

We then move on to the HLZ approach. Just like the case of tensor products of modules over commutative rings, they define fusion via a universal property which we cover in \cref{sec:mathdef}. Here, we introduce and explain the concept of intertwining maps.  These maps are a rigorous version of the physicists' rather general notion of a field and are central to the universal property. Of course, this definition of fusion is quite abstract and it is not obvious how to go about actually computing it.  One therefore needs a concrete model of this universal gadget to realise the fusion product concretely. We begin to explain this in \cref{sec:models}, working with a model built out of the vector space tensor product of modules, and we explain its shortcomings in terms of convergence issues. By the end of this \lcnamecref{sec:models}, we will have understood Gaberdiel's definition of the fusion product in terms of the universal property of HLZ.

In order to cure the divergences of this model, one can pass to the dual of the vector space tensor product.  This is the key to the rigorous formalism developed by HLZ, as we explain in \cref{sec:doubledual}.
Here we shall present a succinct exposition of the crucial ``$P(z)$-compatibility condition'' to build fusion products, although we shall use the complex variable $w$ instead of $z$ to make the relation with the physicists' methods more transparent.
We end this section by working out what the HLZ formalism means in the example of the ``simplest'' possible vertex operator algebras --- those associated to commutative associative unital algebras.

\cref{sec:virex} is devoted to working out the details, using the HLZ formalism, of a specific example of a fusion product. We choose to fuse a certain highest-weight Virasoro module of central charge $-2$ with itself for historically significant reasons, see \cite{GurLog93}. We then introduce, in \cref{sec:algngk}, some basic features of the NGK fusion algorithm by dualising the algorithm we followed for HLZ.  This is also illustrated with the same example, mostly to make the parallel methodology manifest but also to explain the important roles played by Nahm's ``special subspaces'' (rediscovered by mathematicians as $C_1$-spaces) and ``spurious states''. We mention that the NGK algorithm has not only been successfully applied to a multitude of Virasoro fusion products \cite{GabInd96,EbeVir06,RidPer07,RidLog07,RidPer08,RasCla11,MorKac15}, but has also been used to calculate fusion rules for the $N=1$ superconformal algebras \cite{CanRasRidNS15,CanRidRam16}, triplet algebras \cite{GabRat96,GabFus09,WooFus10,TsuTen13}, fractional-level \wzw{} models \cite{GabFus01,RidFus10,CreRel11} and bosonic ghosts \cite{RidBos14}.

We shall close this paper in \cref{sec:otherapproaches} by providing a quick summary of some of the other approaches to fusion, highlighting their similarities and differences to the NGK and HLZ approaches.  Our lack of expertise with these alternative methods means that this \lcnamecref{sec:otherapproaches} is far less detailed and we ask the reader for forgiveness in case of unintentional omissions.  In any case, we hope that the literature we do mention will be of some use to those interested in studying these alternative approaches and will perhaps inspire future comparisons between them and NGK or HLZ.

\subsection*{Acknowledgements}
This paper was made possible by an Endeavour Research Fellowship, ID 6127\_2017, awarded to SK by the Australian Government's Department of Education and Training.
SK wishes to express sincere gratitude towards the School of Mathematics and Statistics at the University of Melbourne, where this project was undertaken, for their generous hospitality. SK is presently supported by a start-up grant provided by University of Denver.
DR's research is supported by the Australian Research Council Discovery Project DP160101520 and the Australian Research Council Centre of Excellence for Mathematical and Statistical Frontiers CE140100049.

It is our privilege to thank our fellow ``fusion club'' members Arun Ram and Kazuya Kawasetsu for the many hours that we spent together working through the details of the approaches of NGK, HLZ, Kazhdan-Lusztig and Miyamoto.  We also thank Thomas Creutzig, Hubert Saleur and Simon Wood for encouraging us to complete this article when time was lacking and deadlines were passing.  We similarly thank Dra\v{z}en Adamovi\'{c} and Paolo Papi for generous amounts of leeway in regard to this last point.

\section{Some notation and conventions}
\label{sec:language}

Our aim is to explain some of
the deep constructions of Huang-Lepowsky-Zhang (HLZ) \cite{HLZ}, in an manner accessible to physicists, while providing
mathematicians with an opportunity to
grasp the ideas of Nahm and Gaberdiel-Kausch (NGK) \cite{NahQua94,GabInd96}.  The former provides a path to make fusion rigorous and the latter shows how to algorithmically implement the computation and identification of fusion products.  There is, of course, significant overlap, but the language used by the authors (being mathematicians and physicists, respectfully) differs markedly.
Even before we present precise definitions, which we shall do in due course,
we feel that it will be useful to address this divide by providing a short
dictionary of terms and notations that shall be used below.

Let $V$ be a \voa{}. 
By the state-field correspondence, there is a bijection between the space of fields $v(z)$ and the space of states $v\in V$.  This state space is, of course, the \voa{}.  As a $V$-module, it is also commonly referred to as the vacuum module.
In the mathematical literature, the field $v(z)$ is typically denoted by $Y(v,z)$.  We shall use both notations interchangeably. While the axiomatic treatment presented in, for example, \cite{FLM, KacVer96,FreVer01,LeLi}
often regards $z$ as a formal variable,
we shall adhere to the physicists' convention of always working with a complex variable $z$, unless otherwise indicated.

In any \voa{}, there are two distinguished states: the vacuum, which we shall denote by $\vacuum$, and the conformal vector, which we shall denote by $\confvec$.  The former corresponds to the identity field $1$, while the latter yields the energy-momentum (or stress-energy) tensor $T(z)$ whose Fourier modes (see \eqref{eq:Tmodes} below) are identified with the Virasoro generators $L(n)$, $n \in \ZZ$.  The most common notation, particularly in the physics literature, for the vacuum state is of course $\lvert0\rangle$.  However, we shall avoid using ``bra-ket notation'' entirely, reserving $\inner{\cdot}{\cdot}$ to denote the usual pairing (whose output is a complex number) between a vector space and its dual.

It is important to note that physicists also use the term ``field'' to denote objects $\psi(z)$ that ``correspond'' (in the same sense as that of the state-field correspondence above) to the elements $\psi$ of a given $V$-module $M$.  More precisely, the fields $\psi(z)$ should be regarded as the chiral, or holomorphic, part of the ``bulk fields'' of the conformal field theory, the latter being objects $\psi(z,\overline{z})$ that correspond to linear combinations of elements in certain vector space tensor products $M \otimes \overline{M}$ of $V$-modules.\footnote{In general, one might instead have tensor products $M \otimes \overline{M}$, where $M$ is a $V$-module and $\overline{M}$ is a module over another \voa{}, $\overline{V}$ say.  In either case, $M$ is responsible for the $z$-dependence of the field $\psi(z,\overline{z})$ while the $\overline{z}$-dependence comes from $\overline{M}$.}  It is also important to note that physicists do not explain what (bulk) fields actually are --- fields are fundamental objects in quantum field theory and so need not be explained as long as one can calculate with them.  The mathematicians' approach to (the holomorphic part of these) fields will be discussed below in \cref{def:intwmap}.  We are only interested in such holomorphic parts $\psi(z)$ here.

Another major notational difference between the physics and mathematics literature concerns
the Fourier coefficients of the fields.
In mathematics, the modes are frequently
given as follows ($\mathfrak{m}$ stands for mathematics):
\begin{align} \label{eq:FourierConvention}
Y(v,z) = \sum_{n\in\ZZ}v_n^{(\mathfrak{m})} z^{-n-1}.
\end{align}
This has some advantages.  First, it applies uniformly to all fields in the \voa{}.  Second, the shift by $-1$ in the exponent of $z$ makes residue calculations particularly easy.  Unfortunately, it also has some computational disadvantages, chief among which is that mode indices are not conserved.  For example, derivatives and normally ordered products of fields have modes satisfying
\begin{equation} \label{eq:mathmoding}
	(\pd v)_n^{(\mathfrak{m})} = -n v_{n-1}^{(\mathfrak{m})} \qquad \text{and} \qquad
	\normord{vv'}_n^{(\mathfrak{m})} = \sum_{r \le -1} v_r^{(\mathfrak{m})} v_{n-1-r}^{(\mathfrak{m})} + \sum_{r>-1} v_{n-1-r}^{(\mathfrak{m})} v_r^{(\mathfrak{m})},
\end{equation}
respectively.  This moreover leads to a conflict with the established convention for the Virasoro generators and so mathematicians almost always modify \eqref{eq:FourierConvention} for the energy-momentum tensor:
\begin{align} \label{eq:Tmodes}
Y(\confvec,z)
= \sum_{n\in\ZZ}L(n) z^{-n-2}.
\end{align}

The expansion \eqref{eq:Tmodes} is in fact an example of the convention used universally by physicists.
Given a state $v$ of $L(0)$-eigenvalue $h_v$, called the conformal dimension or conformal weight of $v$, the corresponding field is expanded
as ($\mathfrak{p}$ stands for physics)
\begin{align}
v(z) = \sum_{n\in\ZZ {-h_v}}v_n^{(\mathfrak{p})} z^{-n-h_v}.
\end{align}
We therefore have
\begin{align}
v_n^{(\mathfrak{m})} = v_{n+1-h_v}^{(\mathfrak{p})} \qquad \text{and so} \qquad
v_n^{(\mathfrak{p})} = v_{n-1+h_v}^{(\mathfrak{m})}.
\end{align}
The obvious disadvantage is that this only applies to states and fields of definite conformal weight and must be extended by linearity in general.  It also means that residue calculations require a lot of $h_v$ factors.  The main advantage, aside from \eqref{eq:Tmodes} being the natural expansion of $T(z)$, is that mode indices reflect the $L(0)$-grading:
\begin{equation}
	\comm{L(0)}{v_n^{(\mathfrak{p})}} = -n \, v_n^{(\mathfrak{p})}.
\end{equation}
The physicists' analogues of \eqref{eq:mathmoding} illustrate the consequent conservation of mode indices as well as the omnipresent $h_v$ factors:
\begin{equation}
	(\partial v)_n^{(\mathfrak{p})} = -(n+h_v) v_n^{(\mathfrak{p})} \qquad \text{and} \qquad
	\normord{vv'}_n^{(\mathfrak{p})} = \sum_{r \le -h_v} v_r^{(\mathfrak{p})} v_{n-r}^{(\mathfrak{p})} + \sum_{r>-h_v} v_{n-r}^{(\mathfrak{p})} v_r^{(\mathfrak{p})}.
\end{equation}
We also mention that the ``zero modes'' that play such an important role in classifying modules over \voas{} \cite{FeiAnn92,FreVer92,ZhuMod96}, are only modes with index $0$ if we employ the physicists' convention.  Despite these computational advantages, our study of fusion will require a number of residue computations that turn out to be much cleaner with the mathematics convention.  The default convention we use is therefore that of \eqref{eq:FourierConvention}.  We shall drop the superscript $(\mathfrak{m})$ in what follows for brevity: $v_n \equiv v_n^{(\mathfrak{m})}$.

\section{Fusion: The physicists' approach} \label{sec:NGK}

Fusion was originally introduced by physicists studying rational \cft{} in order to keep track of which primary fields appeared in the \ope{} of two primary fields.  We recall that a primary field is one that corresponds to a (Virasoro) \hwv{} under the state-field correspondence (extended to include non-vacuum modules).  The standard means for computing the fusion of these two primary fields was then to calculate every correlation function of three primary fields in which two were the primaries being fused.  If the result was found to be non-zero, then the fusion would include (the conjugate of) the third primary field.  This led to \emph{fusion rules} that were expressed as follows:
\begin{equation} \label{eq:fusingprimaries}
	\psi_i \fuse \psi_j = \sum_k \fuscoeff{i}{j}{k} \psi_k.
\end{equation}
Here, the $\psi_k$ are the primary fields, indexed by some (discrete) set, $\fuse$ is the \emph{fusion product}, and the $\fuscoeff{i}{j}{k} \in \ZZ_{\ge 0}$ are the \emph{fusion coefficients} or \emph{fusion multiplicities}.  The meaning of these coefficients, when not $0$ or $1$, is somewhat obscure in this framework.  We shall now reinterpret fusion rules in a manner that makes this meaning transparent.

In rational theories, conformal invariance allows one to compute correlation functions involving non-primary fields from primary ones.  Physicists would therefore speak of the fusion of ``conformal families''.  It is now not difficult to realise that the physicists' notion of a conformal family is morally identical to the mathematician's notion of a \hwm{}.  Inevitably, the idea arose that the fusion product should actually be regarded as a product of modules.  More precisely, any two modules $M$ and $N$ over a given \voa{} $V$ should admit a fusion product, which we shall denote by $M \fus{} N$ to avoid confusion with direct products, that is also (naturally) a $V$-module.  In rational \cft{}, the original fusion product \eqref{eq:fusingprimaries} of primary fields is thus upgraded to the following fusion product of irreducible \hw{} $V$-modules $M_k$:
\begin{equation} \label{eq:fusingmodules}
	M_i \fus{} M_j \cong \bigoplus_k \fuscoeff{i}{j}{k} M_k.
\end{equation}
We remark that with this formulation, there is no longer any reason to require that the theory be rational nor that the $M_k$ be irreducible and \hw{}.\footnote{Note that a non-rational theory may also possess fields that are not generated from primary fields.  The original approach to computing fusion from primary correlators will therefore produce incorrect fusion rules in general.  Unfortunately, this approach is still widely employed, without comment, in the non-rational physics literature.}

As an aside, we mention that mathematicians have since appropriated some of the terminology invented by physicists for fusion, but unfortunately use it in a different, and potentially confusing, way.  In particular, the non-negative integers $\fuscoeff{i}{j}{k}$ have come to be known as fusion rules in the mathematical literature.  In what follows, we shall eschew this and always follow the nomenclature introduced by physicists.  While there is a clear case to be made for respecting original terminology, which is anyway universally used in the physics literature, we also mention that using ``fusion rule'' for the explicit decomposition \eqref{eq:fusingmodules} of a fusion product into (indecomposable) $V$-modules accords well with the universally accepted usage of the group-theoretic term ``branching rule'' for the corresponding decompositions of restrictions of modules.

In any case, once one has decided to view fusion as a product on an appropriate category of $V$-modules, instead of in terms of correlators and \opes{} of primary fields, several questions naturally arise:
\begin{enumerate}
	\item What is the precise definition of fusion?
	\item How can one actually compute it?
	\item Is the result again a $V$-module? \label{it:fusiongivesV-mods}
	\item Which modules can actually be fused?
\end{enumerate}
The last two questions, which are actually about the module category, is perhaps outside the physicists' remit for clearly one must be able to fuse modules appearing in a conformal field theory in order to have a consistent theory.  However, the first two were tackled by Gaberdiel in \cite{GabFus94,GabFus94b} who credits unpublished work of Borcherds and the widely influential work of Moore and Seiberg \cite{MooSeiCFT89} for inspiration.  We review his answers in the rest of this \lcnamecref{sec:NGK}, referring to the original papers for further details.  Our treatment follows \cite[App.~A]{CanRasRidNS15}.

Gaberdiel's algebraic reformulation of fusion begins by considering the following contour integral:
\begin{equation} \label{eq:ngkstart}
	\Res{z}{0,w_1,w_2}{\inner{\psi_3'}{v(z) \psi_1(w_1) \psi_2(w_2) \vacuum} z^n}.
\end{equation}
Here, $\psi_3'$ is an arbitrary vector in the space dual to the states of the \cft{} (see \cref{sec:mathdef} below for a precise definition), $v(z) = \sum_{m \in \ZZ} v_m z^{-m-1}$ is an arbitrary field of the \voa{} $V$ (note mathematics mode convention in force!), $\psi_1(w_1)$ and $\psi_2(w_2)$ are arbitrary (chiral) fields corresponding to states in the $V$-modules $M_1$ and $M_2$, respectively, $n \in \ZZ$ is arbitrary, and the contour integral indicates that the result is the sum of the residues of the integrand at $z=0$, $z=w_1$ and $z=w_2$.  In other words, we consider a simple positively oriented contour that encloses the points $0$, $w_1$ and $w_2$.  When we discuss the mathematical definition of the fields $\psi_i(w_i)$ below, we shall see that they are closely related to certain ``intertwining maps'' evaluated at the states $\psi_i$.  The idea behind considering \eqref{eq:ngkstart} is that this expression defines a natural action of the mode
\begin{equation}
	v_n = \Res{z}{0}{v(z) z^n}
\end{equation}
on the product $\psi_1(w_1) \psi_2(w_2)$ of fields in a correlator and thence on the tensor product states $\psi_1 \otimes \psi_2$ in the vector space tensor product $M_1 \otimes M_2$.  Note that the usual radial ordering prescription of conformal field theory requires us to assume that $\abs{z} > \abs{w_1} > \abs{w_2}$ in \eqref{eq:ngkstart}.

We suppose that each field $\psi_i(w_i)$, $i=1,2$, is mutually local with respect to $v(z)$, meaning that
\begin{equation} \label{eq:ngklocality}
	v(z) \psi_i(w_i) = \psi_i(w_i) v(z), \qquad i=1,2.
\end{equation}
In general, this \lcnamecref{eq:ngklocality} would be modified by adding a coefficient $\mu_i$ on the \rhs{}.  For example, we would have $\mu_i = 1$, if $v(z)$ and $\psi_i(w_i)$ are mutually bosonic, and $\mu_i = -1$ if they are mutually fermionic.  More complicated mutual localities are of course possible (and easily accommodated in the derivation to follow).  For simplicity, we shall assume bosonic statistics ($\mu_i = 1$) throughout.  We also suppose that both $M_1$ and $M_2$ are untwisted as $V$-modules, meaning that the \opes{} of their fields with those of $V$ have trivial monodromy.  In particular, we have
\begin{equation} \label{eq:ngkope}
	v(z) \psi_i(w_i) = \sum_{m \in \ZZ} (v_m \psi_i)(w_i) (z-w_i)^{-m-1}, \qquad i=1,2,
\end{equation}
where $(v_m \psi_i)(w_i)$ denotes the field corresponding to the state $v_m \psi_i \in M_i$.  Note that requiring $M_1$ and $M_2$ to be untwisted excludes, for example, modules from the Ramond sectors of theories with fermions.  Gaberdiel's formalism can, of course, be generalised to accommodate twisted modules though it becomes significantly more unwieldy, see \cite{GabFus97,CanRidRam16}.

Consider the contribution to \eqref{eq:ngkstart} corresponding to the residue at $w_1$.  If we substitute the $i=1$ \ope{} \eqref{eq:ngkope} into this contribution, we (formally) obtain
\begin{subequations}
	\begin{align}
		\Res{z}{w_1}{\inner{\psi_3'}{v(z) \psi_1(w_1) \psi_2(w_2) \vacuum} z^n}
		&= \sum_{m \in \ZZ} \inner{\psi_3'}{(v_m \psi_1)(w_1) \psi_2(w_2) \vacuum} \Res{z}{w_1}{(z-w_1)^{-m-1} z^n} \label{eq:ngkres1a} \\
		&= \sum_{m=0}^{\infty} \binom{n}{m} w_1^{n-m} \inner{\psi_3'}{(v_m \psi_1)(w_1) \psi_2(w_2) \vacuum}. \label{eq:ngkres1b}
	\end{align}
\end{subequations}
To get the contribution from $w_2$, we instead substitute the $i=2$ \ope{}, after first applying \eqref{eq:ngklocality}.  We mention the easily overlooked fact that radial ordering on the \lhs{} requires $|z|>|w_1|$, but re-writing the integral using \eqref{eq:ngklocality} requires $|w_1|>|z|$.  The result is
\begin{equation} \label{eq:ngkres2}
	\Res{z}{w_2}{\inner{\psi_3'}{v(z) \psi_1(w_1) \psi_2(w_2) \vacuum} z^n} = \sum_{m=0}^{\infty} \binom{n}{m} w_2^{n-m} \inner{\psi_3'}{ \psi_1(w_1) (v_m \psi_2)(w_2) \vacuum}.
\end{equation}
If $n \ge 0$, either substitution shows that the residue at $0$ vanishes, hence that \eqref{eq:ngkstart} is the sum of \eqref{eq:ngkres1b} and \eqref{eq:ngkres2}.  Gaberdiel's conclusion is that the arbitrariness of $\psi_3'$ means that we should interpret this sum as the action of $v_n$ on the tensor product state $\psi_1 \otimes \psi_2 \in M_1 \otimes M_2$ corresponding to the product $\psi_1(w_1) \psi_2(w_2)$ under an extended state-field correspondence:
\begin{align} \label{eq:ngkcoproduct+ve}
	\copropts{w_1,w_2}{v_n} (\psi_1 \otimes \psi_2) &= \sum_{m=0}^n \binom{n}{m} \sqbrac[\Big]{w_1^{n-m} (v_m \psi_1) \otimes \psi_2 + w_2^{n-m} \psi_1 \otimes (v_m \psi_2)} \notag \\
	\implies \qquad \copropts{w_1,w_2}{v_n} &= \sum_{m=0}^n \binom{n}{m} \sqbrac[\Big]{w_1^{n-m} (v_m \otimes \wun) + w_2^{n-m} (\wun \otimes v_m)} & \text{(}n &\ge 0\text{).}
\end{align}
Here, $\wun$ denotes the identity operator acting on $M_1$ or $M_2$, as appropriate.  This action first appeared (with $w_2=0$) in work of Moore and Seiberg, see \cite[Eq.~(2.4)]{MooCla89}.

Suppose now that $n<0$.  Then, the contribution to \eqref{eq:ngkstart} from $0$ need not vanish.  To compute it, we have to use one of the two \opes{} \eqref{eq:ngkope}.  If we use that with $i=1$, then the result is the same as in \eqref{eq:ngkres1a} except that the residue is evaluated at $z=0$.  The sum of the contributions from $0$ and $w_1$ may therefore be expressed as a sum over $m \in \ZZ$ of terms proportional to
\begin{equation} \label{eq:ngkres3}
	\Res{z}{0,w_1}{(z-w_1)^{-m-1} z^n} = -\Res{z}{\infty}{(z-w_1)^{-m-1} z^n} = \Res{y}{0}{(1-w_1 y)^{-m-1} y^{m-n-1}} \qquad \text{($y = z^{-1}$),}
\end{equation}
which vanishes for $m>n$.  Adding the $m \le n$ contributions to that from $w_2$, given in \eqref{eq:ngkres2}, we obtain a formula for the action of the $v_n$:
\begin{equation} \label{eq:ngkcoproduct1-ve}
	\coproipts{1}{w_1,w_2}{v_n} = \sum_{m=-\infty}^n \binom{-m-1}{-n-1} (-w_1)^{n-m} (v_m \otimes \wun) + \sum_{m=0}^{\infty} \binom{n}{m} w_2^{n-m} (\wun \otimes v_m) \qquad \text{($n<0$).}
\end{equation}
Note that the upper limit on the first sum may be changed to $-1$ because $\binom{-m-1}{-n-1} = 0$ if $n<m<0$.

In \eqref{eq:ngkcoproduct1-ve}, we have added the label $(1)$ to the action of $v_n$ because, in deriving it, we made a choice to use \eqref{eq:ngkope} with $i=1$ to evaluate the contribution from $0$.  If we had instead chosen to take $i=2$, then we would have deduced a seemingly different action:
\begin{equation} \label{eq:ngkcoproduct2-ve}
	\coproipts{2}{w_1,w_2}{v_n} = \sum_{m=0}^{\infty} \binom{n}{m} w_1^{n-m} (v_m \otimes \wun) + \sum_{m=-\infty}^n \binom{-m-1}{-n-1} (-w_2)^{n-m} (\wun \otimes v_m) \qquad \text{($n<0$).}
\end{equation}
Gaberdiel's definition of the fusion product of $M_1$ and $M_2$ is then the largest quotient of the vector space tensor product on which these two actions agree.  We shall make this manifest shortly after discussing the role of the insertion points $w_1$ and $w_2$.

The symbol $\coprosymb$ was chosen by Gaberdiel to indicate these actions because they define coproducts, albeit ones that depend on the two points $w_1$ and $w_2$.  Indeed, he proved coassociativity\footnote{To be precise, Gaberdiel showed how to prove it for Virasoro \voas{} in \cite[App.~B]{GabFus94}.  Coassociativity in general is stated to follow similarly in \cite[Sec.~2]{GabFus94b}.} in the form
\begin{equation}
	(\coprono{i}{w_1-w,w_2-w} \otimes \wun) \circ \coprono{i}{w,w_3} = (\wun \otimes \coprono{i}{w_2-w,w_3-w}) \circ \coprono{i}{w_1,w}, \qquad \text{($i=1,2$).}
\end{equation}
He moreover showed that conjugating with translation and dilation operators allows one to replace the points $w_1$ and $w_2$ in these coproducts with any two distinct points on the Riemann sphere without changing the equivalence class of the coproduct actions.

Doing so has an immediate practical advantage.  While the modes $v_m$ of any field of $V$ must annihilate an arbitrary fixed state, when $m$ is sufficiently large, there is no such requirement for $m$ sufficiently small.  It follows that the second infinite sum of \eqref{eq:ngkcoproduct1-ve} and the first infinite sum of \eqref{eq:ngkcoproduct2-ve} are both truncated to finite sums when acting on any state.  Unfortunately, no such truncation occurs for the first infinite sum of \eqref{eq:ngkcoproduct1-ve} or the second infinite sum of \eqref{eq:ngkcoproduct2-ve}.  These coproduct actions therefore take elements of $M_1 \otimes M_2$ to some, as yet uncharacterised, completion of this vector space tensor product.  However, we may avoid having to introduce this completion by choosing $w_1 = 0$ in \eqref{eq:ngkcoproduct1-ve} and $w_2 = 0$ in \eqref{eq:ngkcoproduct2-ve}:
\begin{subequations} \label{eq:ngkcoproduct-ve}
	\begin{align}
		\coproipts{1}{0,w_2}{v_n} = (v_n \otimes \wun) + \sum_{m=0}^{\infty} \binom{n}{m} w_2^{n-m} (\wun \otimes v_m) \qquad \text{($n<0$),} \label{eq:ngkspeccoproduct1-ve} \\
		\coproipts{2}{w_1,0}{v_n} = \sum_{m=0}^{\infty} \binom{n}{m} w_1^{n-m} (v_m \otimes \wun) + (\wun \otimes v_n) \qquad \text{($n<0$).} \label{eq:ngkspeccoproduct2-ve}
	\end{align}
\end{subequations}
These coproduct formulae thus give well defined actions on $M_1 \otimes M_2$.  Moreover, they now agree with the corresponding specialisations of the $n \ge 0$ formulae of \eqref{eq:ngkcoproduct+ve}.  Note that the contribution from $0$ to the $n<0$ formula for $\coprono{i}{w_1,w_2}$ was derived by inserting an \ope{} at $w_i$.  We could have therefore insisted from the outset that $w_i$ be required to be close to $0$.  Setting $w_i$ to $0$ at the conclusion is thus very natural.

\begin{example}
	We illustrate a few coproduct formulae in order to appreciate their complexity.  For brevity, we use \eqref{eq:ngkcoproduct+ve}, with $w_1 = 1$ and $w_2 = 0$, and \eqref{eq:ngkspeccoproduct2-ve}, with $w_1 = 1$:
	\begin{subequations}
		\begin{align}
			\copropts{1,0}{v_2} &= (v_0 \otimes \wun) + 2 (v_1 \otimes \wun) + (v_2 \otimes \wun) + (\wun \otimes v_2), \label{eqn:D10phi2} \\
			\copropts{1,0}{v_1} &= (v_0 \otimes \wun) + (v_1 \otimes \wun) + (\wun \otimes v_1),\label{eqn:D10phi1} \\
			\copropts{1,0}{v_0} &= (v_0 \otimes \wun) + (\wun \otimes v_0), \label{eqn:D10phi0}\\
			\coproipts{2}{1,0}{v_{-1}} &= (v_0 \otimes \wun) - (v_1 \otimes \wun) + (v_2 \otimes \wun) - \cdots + (\wun \otimes v_{-1}), \label{eqn:D102phim1} \\
			\coproipts{2}{1,0}{v_{-2}} &= (v_0 \otimes \wun) - 2 (v_1 \otimes \wun) + 3 (v_2 \otimes \wun) - \cdots + (\wun \otimes v_{-2}). \label{eqn:D102phim2}
		\end{align}
	\end{subequations}
	Note that the coproduct formula for the (mathematics convention!) ``zero modes'' matches that used for Lie algebra representations.  These ``zero modes'' include the Virasoro mode $L(-1) = \confvec_0$.
\end{example}

We can now make Gaberdiel's definition of the fusion product of $M_1$ and $M_2$ precise.  Taking $w_1=1$ and $w_2=-1$ in \eqref{eq:ngkcoproduct-ve}, for maximum brevity, the insertion points of the two coproducts are related by a rigid translation.  Imposing the equality of the two coproduct actions therefore amounts to setting
\begin{equation} \label{eq:setcoprodequal}
	\coproipts{1}{0,-1}{v_n} = \coproipts{2}{0,-1}{v_n} = \coproipts{2}{1,0}{\ee^{L(-1)} v_n \ee^{-L(-1)}},
\end{equation}
for all fields $v(z)$ and all $n<0$.  The fusion product is therefore defined to be the vector space
\begin{equation} \label{eq:Gabdef}
	M_1 \fus{} M_2 = \frac{M_1 \otimes M_2}{\ideal{\brac[\big]{\coproipts{1}{0,-1}{v_n} - \coproipts{2}{1,0}{\ee^{L(-1)} v_n \ee^{-L(-1)}}} (M_1 \otimes M_2)}},
\end{equation}
equipped with the action of $V$ defined by the coproduct formulae \eqref{eq:ngkcoproduct+ve} and either \eqref{eq:ngkcoproduct1-ve} or \eqref{eq:ngkcoproduct2-ve}.  Here, the quotient is by the sum of the images for all fields $v(z)$ of $V$ and all $n<0$.  As mentioned above, the point is that the fusion product is constructed to be the largest quotient of the vector space tensor product on which the natural $V$-action, derived from \opes{} and locality, is well defined.  This foreshadows the idea that the definition should be reinterpreted in terms of a universality property.

Unfortunately, fusion is not all tea and biscuits.  While the issue of having to specify a completion of $M_1 \otimes M_2$ was neatly sidestepped by choosing insertions points carefully, a nastier problem now rears its ugly head: the translations required to compare the two coproducts.  Specifically, the Lie bracket $\comm{L(-1)}{v_n} = -n v_{n-1}$ implies that \eqref{eq:setcoprodequal} may be expanded into
\begin{equation} \label{eq:coprodidentity}
	\coproipts{1}{0,-1}{v_n} = \sum_{m=0}^{\infty} (-1)^m \binom{n}{m} \coproipts{2}{1,0}{v_{n-m}}.
\end{equation}
Inserting the coproduct formula on the \rhs{}, we find that the sum does not converge for $n\le-1$, not even when acting on $M_1 \otimes M_2$ and taking a completion:
\begin{multline} \label{eq:doesnotconverge}
	\coproipts{1}{0,-1}{v_n} = \sum_{m=0}^{\infty} (-1)^m \binom{n}{m} (v_0 \otimes \wun) + \sum_{m=0}^{\infty} (-1)^m \binom{n}{m} (n-m) (v_1 \otimes \wun) \\
	+ \sum_{m=0}^{\infty} (-1)^m \binom{n}{m} \binom{n-m}{2} (v_2 \otimes \wun) + \cdots + \sum_{m=0}^{\infty} (-1)^m \binom{n}{m} (\wun \otimes v_{n-m}).
\end{multline}

Gaberdiel was certainly aware that issues like this arose whenever one combined coproducts and translations, see \cite[Eq.~(2.14)]{GabFus94} and the subsequent discussion.  However, he did not offer any solutions.  Nahm, in his seminal paper on the definition and computation of fusion \cite{NahQua94}, noted that such convergence issues may be resolved by either working with dual modules or by redefining the fusion product as a projective limit of finite-dimensional truncations of $M_1 \otimes M_2$, see \cref{sec:algngk}.\footnote{This inverse limit approach to fusion has also reappeared in the work of Miyamoto \cite{MiyC114} and Tsuchiya and Wood \cite{TsuTen13}.}  At around the same time, Huang and Lepowsky \cite{HuaTheI95,HL-vtc} were likewise formulating their definition of fusion, which we shall review shortly, using dual spaces.

However, Nahm's truncation idea turned out to be much more interesting to physicists as it was subsequently generalised by Gaberdiel and Kausch \cite{GabInd96} and used to form the basis of a practical algorithm to explicitly construct (truncations of) fusion products.  We shall discuss this algorithm, now known as the \ngk{} fusion algorithm, in \cref{sec:algngk}.  First, however, we shall compare Gaberdiel's definition of fusion with the dual definition of Huang and Lepowsky.

\section{The mathematician's approach I: A universal definition}
\label{sec:mathdef}

There are many essentially equivalent definitions of \voas{}, which the reader may find in references such as \cite{FLM,KacVer96,FreVer01}.  The main axiom defining modules for a given \voa{} $V$ is what is called the Jacobi identity in \cite{LeLi}.  We will be interested in $V$-modules $M$ which are graded
by the zero mode $L(0)$ of the Virasoro field.  More precisely,
we require that $M$ be spanned, and thus graded, by its \emph{generalised} $L(0)$-eigenspaces --- in particular, we allow non-diagonalisable actions of $L(0)$.  We denote the generalised eigenspace
of $M$ corresponding to the eigenvalue $h\in\CC$ by $M_{[h]}$; this eigenspace decomposition is called the \emph{conformal grading}.  We shall insist, with a caveat to be discussed below, that each $M_{[h]}$ is finite-dimensional.  Then, restricting $L(0)$ to $M_{[h]}$ results in finite-rank Jordan blocks.

One is also required to truncate the conformal grading from below, in some sense, because the action of a field on a state of a module should be expressed as a Laurent series with poles, but no essential singularities.  This translates into the following requirement: For every $v \in V$ and $\psi \in M$, we have $v_n \psi = 0$ for all sufficiently large $n$.  However, one frequently finds stronger conditions being
assumed in the mathematical literature, for example
that for any given $h\in\CC$, the spaces $M_{[h-n]}$ are zero for all sufficiently large integers $n$.  This stronger condition is met when $V$ is $C_2$-cofinite (see \cite{AbeRat04}), but can be inappropriate in general.  In particular, the ``staggered/logarithmic'' modules of the admissible-level affine \cite{GabFus01,AdaLat09,RidFus10} and bosonic ghost \cite{RidVer14} \voas{} all fail to meet this condition.

It may also be highly beneficial, or indeed necessary, to introduce additional gradings by other zero modes. The most frequent (and natural) occurrence of this arises when $V$ includes a Heisenberg \vosa{} whose zero modes act semisimply on an appropriate class of $V$-modules.  Such additional gradings are used, even when $V$ is rational, to refine characters so that they may be used to distinguish inequivalent irreducible $V$-modules.  When $V$ is not $C_2$-cofinite, they may be required for characters to even be defined because the generalised $L(0)$-eigenspaces may be infinite-dimensional.  Insisting on finite-dimensional homogeneous spaces with respect to the conformal and additional gradings then saves the notion of character while also preserving the finite-rank property for Jordan blocks.
The theorems of HLZ \cite{HLZ} are designed to handle such additional gradings.
However, for the sake of simplicity, we will not emphasise this level of generality in what follows.

To proceed with the rigorous definitions, we first say what we mean by the \emph{restricted dual} $M'$ of a graded $V$-module $M$.  As a vector space, the definition is straightforward: one merely takes $M'$ to be the direct sum of the duals of the generalised $L(0)$-eigenspaces.  In symbols,
\begin{equation} \label{eq:defvsdual}
	M = \bigoplus_{h \in \CC} M_{[h]} \qquad \implies \qquad M' = \bigoplus_{h \in \CC} M_{[h]}^*.
\end{equation}
Here, ${}^*$ denotes the ordinary vector space dual.\footnote{Because we assume that the homogeneous subspaces $M_{[h]}$ are all finite-dimensional, we may safely ignore all questions regarding the topological nature of these duals.}  This generalises in the obvious way when there are additional grading in force.  In \cref{sec:doubledual}, we shall refine this notion by endowing the restricted dual $M'$ with the structure of a $V$-module.

We shall also need to introduce the notion
of a suitable completion of a module. Generically, the field $Y(v,z)$ (written in
mathematical notation), with $z$ being a non-zero complex parameter and $v \in V$, acts
on $\psi\in M$ to give
an infinite sum of elements in $M$. To accommodate for this (and other analogous infinite sums), we will often work with a completion $\overline{M}$ of a module, in which the direct sum of the generalised $L(0)$-eigenspaces of $M$ is replaced by the direct product.

We now come to the definition of an intertwining map, the objects that form the backbone of the tensor product theory of HLZ.
In mathematics, tensor products are often defined abstractly using ``universal properties'' before proving that the (sometimes) obvious construction of the product satisfies these properties. For example, this is how tensor products of vector spaces are defined in many mathematical textbooks. The main advantages of this universal approach include capturing the uniqueness properties of the tensor product construction as well as its relation to other algebraic and categorical gadgets. In the setup of HLZ \cite{HLZ}, the fusion product introduced by physicists is recast as a universal tensor product object with respect to intertwining maps, as we now explain. A closely related notion, namely that of \emph{intertwining operators}, was introduced much earlier in \cite{FHL} (see \cite{MilWea02} for \emph{logarithmic} intertwining operators).  These are formal variable cousins of the intertwining maps that we consider here.

\begin{dfn} \label{def:intwmap}
	Fix $w\in\CC^\times$.
	Given $V$-modules $M_1$, $M_2$ and $M_3$, a \emph{$P(w)$-intertwining map of type $\intwtype{M_3}{M_1}{M_2}$} is a bilinear map $I\colon M_1\otimes M_2\rightarrow \overline{M_3}$ that satisfies
	the following properties:
	\begin{enumerate}
		\item For any $\psi_1\in M_1$ and $\psi_2\in M_2$, $\pi_h(I[\psi_1\otimes \psi_2])=0$ for all $h\ll0$, where $\pi_h$ denotes the projection onto the generalised eigenspace $(M_3)_{[h]}$ of $L(0)$-eigenvalue $h$.\footnote{This requirement will clearly need refining when it is necessary (or desirable) to include additional gradings on the $V$-modules.}
		\item For any $\psi_1\in M_1$, $\psi_2\in M_2$ and $\psi_3'\in M_3'$, the series defined by
		\begin{equation} \label{eq:threecorrelators}
			\inner*{\psi_3'}{Y_3(v,z)I[\psi_1\otimes \psi_2]},\quad
			\inner*{\psi_3'}{I[Y_1(v,z-w)\psi_1\otimes \psi_2]}\quad \text{and} \quad
			\inner*{\psi_3'}{I[\psi_1\otimes Y_2(v,z)\psi_2]}
		\end{equation}
		are absolutely convergent in the regions $\abs{z}>\abs{w}>0$, $\abs{w}>\abs{z-w}>0$ and $\abs{w}>\abs{z}>0$, respectively.\footnote{In the physics literature, it is customary to take $z$ and $w$ to be the first and second insertion points, respectively, in an \ope{}.  In line with this convention, quantities like \eqref{eq:threecorrelators} naturally lead us to speak of $P(w)$-intertwining maps as opposed to the $P(z)$-intertwining maps that are ubiquitous in the mathematics literature.} (The subscript under each $Y$ indicates the module being acted upon.)
		\item Given any $f(t) \in R_{P(w)} = \CC[t,t^{-1},(t-w)^{-1}]$, the field of rational functions whose poles lie in some subset of $\set{0,w,\infty}$, we have the \emph{Cauchy-Jacobi identity}:
		\begin{align}
			&\Res{z}{0,w}{f(z)\inner*{\psi_3'}{Y_3(v,z)I[\psi_1\otimes \psi_2]}}\notag\\
			&\mspace{40mu}
			=\Res{z}{w}{f(z)\inner*{\psi_3'}{I[Y_1(v,z-w)\psi_1\otimes \psi_2]}}
			+\Res{z}{0}{f(z)\inner*{\psi_3'}{I[\psi_1\otimes Y_2(v,z)\psi_2]}}.
			\label{eqn:IntwMapJacobi}
		\end{align}
		Here, as in \cref{sec:NGK}, the subscript on each integrals indicate the points in $\set{0,w,\infty}$ that must be enclosed by the simple positively oriented contours.
	\end{enumerate}
\end{dfn}

The nomenclature ``$P(w)$'' may look strange, but it emphasises the fact that we are working on
the Riemann sphere with three punctures at $0$, $w$ and $\infty$.  The first two are designated as being ``incoming'', or positively oriented, while the $\infty$ puncture is ``outgoing'', or negatively oriented. We also have a preferred choice of local coordinates around the punctures:
about $0$, we take $z\mapsto z$; about $w$, we take $z\mapsto z-w$; and about $\infty$, we take $z \mapsto z^{-1}$.  The space $R_{P(w)}$ introduced
above is precisely the field of rational functions on this punctured sphere. Below, we shall expand these functions as Laurent
series centred at the punctures, implicitly using the provided local coordinates.  We mention that punctured
spheres are ubiquitous in mathematical approaches to fusion (and conformal field theory in general), see for example Huang's book \cite{HuaBook97} and the seminal work of Kazhdan and Lusztig \cite{KL1,KL2,KL3,KL4}.

Let us also mention that the sewing of such spheres --- outgoing punctures to incoming punctures, respecting the local coordinates --- plays a central role in building the rest of the tensor category structure, most importantly the associativity morphisms. We shall not discuss this further since our focus is only on the tensor (fusion) product itself.
Interested readers may consult \cite{HL-vtc,HuaBook97} for further details.

We now connect the notion of intertwining maps to the objects that were introduced in \cref{sec:NGK}. The entity $\psi_1(w)\psi_2$, in the notation of that \lcnamecref{sec:NGK}, is essentially a $P(w)$-intertwining map,
evaluated on $\psi_1\otimes \psi_2$.
This raises two immediate questions:
\begin{enumerate}
	\item As per the definition, an intertwining map \emph{comes equipped} with a fixed choice of $w$, but in a field such as $\psi_1(w)$, we are free to let $w$ vary. How do we reconcile this?
	\item In the physics literature, the action of the field $\psi_1(w)$, with $\psi_1 \in M_1$, on the state $\psi_2 \in M_2$ may be determined from the corresponding \ope{}, by applying the (generalised) state-field correspondence.  The result of this application is therefore in (the completion of) the fusion product of $M_1$ and $M_2$.  How then does the fusion product relate to the target spaces $M_3$ of the corresponding intertwining maps?
\end{enumerate}
For the first question, it is possible to pass from a $P(w)$-intertwining map (in which $w$ is fixed!) to a formal variable intertwining operator.
This is interesting in itself, and is explained in \cite{HLZ}.  Once this is done,
any other non-zero complex number may be substituted in place of the formal variable. This must be done with great care --- it is necessary to choose branches because the series expansion of the intertwining operator will usually involve non-integer powers and logarithms.
The second question is likewise very interesting.  It leads us to a definition of the fusion product in terms of a universal
tensor product module, $M_1\fus{}M_2$, which we shall define below.

First, however, note that an intertwining map is defined on the whole of the module $M_1 \otimes M_2$, and not just on their ``top spaces'' or primary vectors (or any other ``special'' subspaces).  Nevertheless, a fact that will prove crucial later is that
an intertwining map is often completely determined by its definition on certain
subspaces and/or quotients.  This is quite delicate however.
Given a map on appropriate subspaces and/or quotients, it may require a great deal of effort to prove that it extends
to an intertwining map on the entire module, if at all.

\begin{exam}\label{exam:HeisenbergIntwMap}
	Consider the rank-$1$ Heisenberg vertex operator algebra, known to physicists as the free boson on a one-dimensional non-compact spacetime.  (All of this generalises naturally to higher ranks and dimensions.)
	Its irreducible \hwms{}, known as Fock spaces, are parametrised by complex numbers: for each $\lambda\in\CC$, we have a Fock space
	$F_\lambda$, generated by a \hwv{} $f_\lambda$.
	Fix $w\in\CC^\times$. It can be proved easily that there are no $P(w)$-intertwining maps of type $\intwtype{F_{\nu}}{F_\lambda}{F_{\mu}}$ unless $\nu=\lambda+\mu$. (In the language of physics, this is conservation of momentum.) We leave it as an exercise for the reader to demonstrate, as in \cref{sec:virex}, to prove that there is at most one
	$P(w)$-intertwining map (up to scalar multiples) of the type $\intwtype{F_{\lambda+\mu}}{F_\lambda}{F_{\mu}}$. There indeed exists a non-zero $P(w)$-intertwining map. Up to normalisation, it acts as
	\begin{align}
	I_{\lambda,\mu}[f_\lambda\otimes f_\mu]= w^{\lambda\mu }f_{\lambda+\mu}+\cdots \in \overline{F_{\lambda+\mu}},
	\label{eqn:HeisenbergIntwMap}
	\end{align}
	where the ellipses indicate terms involving factors $w^{\lambda\mu+n}$, with $n$ a positive integer.
	Physicists will indeed recognise this in terms of the \ope{} of the primary fields corresponding to $f_{\lambda}$ (at $w$) and $f_{\mu}$ (at $0$).\footnote{We remark that it is these primary fields (and only these primary fields) that are called \emph{vertex operators} in the physics literature.  In the setting of (non-compact) free bosons, they are therefore not fields of the Heisenberg \voa{}.  The term \emph{\voa{}} itself presumably arose in the mathematical literature because early work concentrated on examples related to lattices (compactified free bosons) in which certain vertex operators are promoted to fields of an extended \voa{}.}  As mentioned above, one still needs to prove that this definition extends to all of $F_\lambda \otimes F_\mu$ (the details may be found in \cite{DonLep93}).  Note that it is important to make use of a specific branch of the logarithm to make sense of terms like $w^{\lambda\mu}$.  However, in this case, different choices only lead to scalar multiples of $I_{\lambda,\mu}$.
\end{exam}

We now give a precise definition of the tensor product by a universal property, as promised above. Let $\categ{C}$ be a category of modules for a \voa{} $V$. We shall define the $P(w)$-tensor product of two modules $M_1, M_2 \in \categ{C}$ as a universal object, denoted by $M_1\fus{P(w)}M_2 \in \categ{C}$, with respect to the $P(w)$-intertwining maps $I\colon M_1\otimes M_2\rightarrow \overline{M_3}$, for all $M_3$ in $\categ{C}$. Comparing with the physics approach of \cref{sec:NGK}, it is clear that this universal approach depends upon the choice of category $\categ{C}$.  In the next two sections, we will separate the categorical and non-categorical constraints.  By focusing only on the latter, we shall produce a rigorous constructive definition of the fusion product of two modules in the formalism of HLZ that can be compared directly with computations performed by physicists.
\begin{dfn} \label{def:univP(w)tp}
	For each $M_1,M_2\in\categ{C}$, the pair $(M_1\fus{P(w)}M_2, \fus{P(w)})$, where $M_1\fus{P(w)}M_2\in \categ{C}$ and $\fus{P(w)}$ is a $P(w)$-intertwining map of type $\intwtype{M_1\fus{P(w)}M_2}{M_1}{M_2}$, is called the \emph{$P(w)$-tensor product} of $M_1$ and $M_2$ if for any $M\in\categ{C}$ and $P(w)$-intertwining map $I$ of type $\intwtype{M}{M_1}{M_2}$, there exists a unique $\categ{C}$-morphism $\eta\colon M_1\fus{P(w)}M_2\to M$ such that
	\begin{align}
		(\overline{\eta}\circ \fus{P(w)})[\psi_1\otimes \psi_2] = I[\psi_1\otimes \psi_2],
	\end{align}
	for all $\psi_1\in M_1$ and $\psi_2\in M_2$.  Here, $\overline{\eta}$ denotes the extension of $\eta$ to a map between the completions of $M_1\fus{P(w)}M_2$ and $M$ (necessary because the image of $I$ is typically in the completion of the target module).  In terms of a diagram, the following should commute:
	\begin{equation} \label{diag:fusuniversal}
		\begin{tikzpicture}[->,>=latex,scale=1.5,baseline={(l.base)}]
			\node (l) at (-1,0) {$M_1\otimes M_2$};
			\node (u) at (1,1) {$\overline{M_1\fus{P(w)} M_2}$};
			\node (d) at (1,-1) {$\overline{M}$};
			\draw (l) -- node[above left] {$\fus{P(w)}$} (u);
			\draw (l) -- node[below left] {$I$} (d);
			\draw (u) -- node[right] {$\overline{\eta}$} (d);
		\end{tikzpicture}
		.
	\end{equation}
\end{dfn}

There is of course the very natural question of whether this definition of $P(w)$-tensor products actually depends on the choice of $w \in \CC^{\times}$.  As expected, HLZ answer this in the negative \cite[Rem.~4.22]{HLZ}.  By abuse of notation, we shall denote $\fus{P(w)}(\psi_1\otimes \psi_2)$ by $\psi_1\fus{P(w)}\psi_2$, for $\psi_1 \in M_1$ and $\psi_2 \in M_2$, keeping in mind that this element is in the completion $\overline{M_1\fus{P(w)}M_2}$ and may not be in the module itself. This subtlety, among others, necessitates the need for analytic arguments throughout \cite{HLZ}.

\begin{exam}
	It is shown in \cite{HLZ} that $V\fus{P(w)}M = M$ for all $V$-modules $M$, with the universal intertwining map $\fus{P(w)}$ of type
	$\intwtype{M}{V}{M}$ being given by $v\fus{P(w)}\psi = Y(v,w)\psi\in\overline{M}$.  On the other hand, we also have $M\fus{P(w)}V = M$, with $\psi\fus{P(w)}v = \ee^{wL(-1)}Y(v,-w)\psi$.  The vacuum module $V$ is thus a unit for the $P(w)$-tensor product, in accordance with expectations for the fusion product.
\end{exam}

\begin{exam}
	Returning to the rank-$1$ Heisenberg \voa{} of \cref{exam:HeisenbergIntwMap}, it can be proved \cite{CKLR} that if $\categ{C}$ is the semisimple category whose objects are finite direct sums of Fock spaces, then one may take $F_\lambda\fus{P(w)}F_\mu  = F_{\lambda+\mu}$, with universal intertwining map $\fus{P(w)}=I_{\lambda,\mu}$ given by \eqref{eqn:HeisenbergIntwMap}.  This also agrees with the well-known ($\categ{C}$-independent) fusion rules known to physicists.
\end{exam}

\section{The mathematician's approach II: A model for tensor products} \label{sec:models}

Universal definitions are all well and good, for some purposes (mostly abstract ones).  But sometimes, one needs an alternative definition that comes with an honest construction.  Mathematicians often refer to such constructions as \emph{models} for the universal definition.  The basic philosophy behind building models satisfying the universal properties is to first take the ``biggest'' candidate possible and then to cut it down by imposing relations arising out of the constraints given by the ``test'' conditions.

With a view towards the vertex-algebraic picture to be presented later, let us review a basic example of a model: the explicit construction of the tensor product $M_1\otimes_A M_2$ of $A$-modules $M_1$ and $M_2$, where $A$ is a commutative associative unital algebra over $\CC$ (say), as a quotient of the vector space tensor product $M_1 \otimes M_2 \equiv M_1 \otimes_{\CC} M_2$.  The universal definition of $\otimes_A$ says that if we are given an $A$-bilinear map $B\colon M_1\otimes M_2 \to M$, where $M$ is an arbitrary $A$-module, then there exists a unique $A$-linear map $f: M_1\otimes_A M_2\rightarrow M$ such that
\begin{equation}
	B(m_1\otimes m_2)=(f\circ\otimes_A)(m_1,m_2)= f(m_1\otimes_A m_2),
\end{equation}
for all $m_1 \in M_1$ and $m_2 \in M_2$.  In other words, the following diagram must commute:
\begin{equation} \label{diag:tensoruniversal}
	\begin{tikzpicture}[->,>=latex,scale=1.5,baseline={(l.base)}]
		\node (l) at (-1,0) {$M_1\otimes M_2$};
		\node (u) at (1,1) {$M_1\otimes_A M_2$};
		\node (d) at (1,-1) {$M$};
		\draw (l) -- node[above left] {$\otimes_A$} (u);
		\draw (l) -- node[below left] {$B$} (d);
		\draw (u) -- node[right] {$f$} (d);
	\end{tikzpicture}
	.
\end{equation}
To construct $M_1 \otimes_A M_2$, we first recall that
the plain old vector space tensor product $M_1\otimes M_2$ is naturally an $A$-module under the action
$a\cdot (m_1\otimes m_2) = m_1 \otimes (a\cdot m_2)$, $a \in A$.\footnote{Here, we choose to act on $m_2$, rather than $m_1$, in order to keep in line with the vertex-algebraic generalisation to follow.}
But, the bilinear map $B$ is constrained by
one more property, namely that $B(a\cdot m_1\otimes m_2)= B(m_1\otimes a\cdot m_2)$. We therefore quotient the $A$-module $M_1\otimes M_2$ by the ideal corresponding to imposing $(a\cdot m_1)\otimes m_2= m_1\otimes (a\cdot m_2)$.  Defining $\coproi{1}{a} = a \otimes \wun$ and $\coproi{2}{a} = \wun \otimes a$, we therefore have the following (hypothetical) model for the universal definition:
\begin{equation}
	M_1 \otimes_A M_2 = \frac{M_1 \otimes M_2}{\ideal{\brac*{\coproi{1}{a} - \coproi{2}{a}}(M_1 \otimes M_2) \st a \in A}}.
\end{equation}
One can of course verify that this definition does satisfy the universal property \eqref{diag:tensoruniversal} and so is indeed a model of the $A$-tensor product.

In all the mathematics and physics literature, this essential idea is behind the explicit constructions of tensor product modules.  In particular, Gaberdiel's original definition \eqref{eq:Gabdef} of fusion is now clearly identified as an attempt to construct a model for the fusion product of two $V$-modules $M_1$ and $M_2$.  However, as \voas{} are highly non-classical objects, one quickly runs into obstacles to making this rigorous.  We also discussed these briefly in \cref{sec:NGK}.  Let us recall them again:
\begin{enumerate}
\item If one tries to work out certain relations analogous to $(a\cdot m_1)\otimes m_2 = m_1\otimes (a\cdot m_2)$ in the vertex-algebraic setting, one finds that they do not converge, even when completions are taken into account, because some coefficients are found to be formally infinite, see \eqref{eq:doesnotconverge}.
\item Unlike the commutative ring case, where $a\cdot (m_1\otimes_A m_2) = m_1\otimes_A (a\cdot m_2)$ trivially satisfies the axioms for the algebra action on a module, it takes much more effort to prove the analogous theorem in the vertex-algebraic setting.  This theorem would answer question \ref{it:fusiongivesV-mods} in the material before \eqref{eq:ngkstart} and we shall say a bit more about this
towards the end of \cref{sec:doubledual}.
\end{enumerate}

In the rest of this \lcnamecref{sec:models}, we shall derive the relations and $V$-action that (morally) should define a model for the universal $P(w)$-tensor product, ignoring these obstacles to making the construction rigorous.  The aim is to draw parallels with Gaberdiel's work, as reviewed in \cref{sec:NGK}, before describing the rigorous formalism developed to this end by HLZ in \cref{sec:doubledual}.

For now, it will be highly beneficial
to detach the $v$ and the $n$ in the (mathematicians') notation for modes
$v_n = v_n^{(\mathfrak{m})}$ by writing them in the form $v\otimes t^n$, where $t$ is some auxiliary formal variable. This is not merely a syntactic vinegar --- it opens up wider possibilities. In particular, we are now working in $V\otimes \CC[t,t^{-1}]$ and we have room to accommodate other regular functions on the $3$-punctured sphere, for instance $v\otimes (t-w)^n$, $n \in \ZZ$, by enlarging further to $V\otimes \CC[t,t^{-1},(t-w)^{-1}]$ or $V \otimes \CC(t)$ (rational functions in $t$) or even $V \otimes \CC((t))$ (Laurent series in $t$).
It will often be convenient to ignore the $\otimes$ sign and write simply $vf(t)$ instead of $v\otimes f(t)$.

Once again, choose $w\in\CC$.  It is convenient to define a translation map
\begin{align}
	T_w \colon \CC(t)\rightarrow \CC(t), \qquad \text{by} \qquad f(t)\mapsto f(t+w),
\end{align}
and two expansion maps
\begin{align}
\iota_+\colon \CC(t)\ira \CC((t)) \qquad \text{and} \qquad \iota_-\colon \CC(t)\ira \CC((t^{-1}))
\end{align}
that expand a given rational function in $t$ as a power series around $t=0$ and $t=\infty$, respectively.

We turn to constructing a model for the fusion product $M_1 \fus{} M_2$, now identified as the $P(w)$-tensor product $M_1 \fus{P(w)} M_2$ defined above.  The relations required to cut $M_1 \otimes M_2$ down to the fusion product naturally arise out of the Cauchy-Jacobi identity \eqref{eqn:IntwMapJacobi} for the intertwining maps, since these are essentially the only constraints we have. Let us therefore analyse this identity closely.

As in \eqref{eqn:IntwMapJacobi}, let $v \in V$, $\psi_1 \in M_1$, $\psi_2 \in M_2$, $\psi_3' \in M_3'$ and let $I$ be a $P(w)$-intertwining map of type $\intwtype{M_3}{M_1}{M_2}$.  Let $f(z)=z^a(z-w)^b$, where $a$ and $b$ are arbitrary integers.
The term on the \lhs{} of \eqref{eqn:IntwMapJacobi} may be expanded formally as
\begin{subequations}
	\begin{align}
		\Res{z}{0,w}{f(z)\inner*{\psi_3'}{Y_3(v,z)I[\psi_1\otimes \psi_2]}}
		&=-\Res{z}{\infty}{z^a(z-w)^b\inner*{\psi_3'}{Y_3(v,z)I[\psi_1\otimes \psi_2]}}\notag\\
		&=\Res{y}{0}{\left(y^{-a-2}(y^{-1}-w)^b\inner*{\psi_3'}{Y_3(v,y^{-1})I[\psi_1\otimes \psi_2]}\right)}\notag\\
		&=\Res{y}{0}{\sum_{m=0}^{\infty} (-1)^m \binom{b}{m} w^m y^{m-a-b-2}
		                   \sum_{n\in\ZZ} \inner*{\psi_3'}{v_n y^{n+1} I[\psi_1\otimes \psi_2]}}\notag\\
		&=\sum_{m=0}^{\infty} (-1)^m \binom{b}{m} w^m \inner*{\psi_3'}{v_{a+b-m}I[\psi_1\otimes \psi_2]}\notag\\
		&=\inner*{\psi_3'}{\brac*{v \, \iota_- \brac[\big]{t^a (t-w)^b}}\cdot I[\psi_1\otimes \psi_2]}.
		\label{eqn:vI}
	\end{align}
	The terms on the \rhs{} of \eqref{eqn:IntwMapJacobi} can similarly be expanded formally as
	\begin{align}
		\Res{z}{w}{f(z) \inner*{\psi_3'}{I[Y_1(v,z-w)\psi_1\otimes \psi_2]}}
		&= \inner*{\psi_3'}{I \sqbrac*{\brac*{v \, \iota_+ T_w \brac[\big]{t^a (t-w)^b} \cdot \psi_1} \otimes \psi_2}} \\
		\text{and} \qquad
		\Res{z}{0}{f(z) \inner*{\psi_3'}{I[\psi_1\otimes Y_2(v,z)\psi_2]}}
		&= \inner*{\psi_3'}{I \sqbrac*{\psi_1\otimes \brac*{v \, \iota_+ \brac[\big]{t^a (t-w)^b} \cdot \psi_2}}}.
	\end{align}
\end{subequations}
The Cauchy-Jacobi identity \eqref{eqn:IntwMapJacobi} may therefore be written in the form
\begin{multline} \label{eqn:jac_complexanalytic}
	\inner[\Big]{\psi_3'}{v \, \iota_-(f(t))\cdot I[\psi_1\otimes\psi_2]} \\
	= \inner[\Big]{\psi_3'}{I\sqbrac[\big]{(v \, \iota_+ T_w(f(t))\cdot \psi_1)\otimes \psi_2}}
	+ \inner[\Big]{\psi_3'}{I \sqbrac[\big]{\psi_1\otimes (v \, \iota_+(f(t))\cdot \psi_2)}} \qquad \text{($f(t) \in R_{P(w)}$).}
\end{multline}

Recall that we are trying to satisfy the universal property described in \eqref{diag:fusuniversal}.
If we were to define an ``action'', denoted by ${\coprosymb_w}$, of $V\otimes R_{P(w)}$
on $M_1\otimes M_2$ by
\begin{align}
	\copropts{w}{v\, f(t)} \cdot (\psi_1\otimes \psi_2)
	= \left(v\, \iota_+T_w \left(f(t)\right)\cdot \psi_1\right)\otimes \psi_2
	+ \psi_1\otimes \left(v\, \iota_+\left(f(t)\right)\cdot \psi_2\right),
	\label{eqn:coprod}
\end{align}
then we could write \eqref{eqn:jac_complexanalytic} succinctly as
\begin{align}
	\inner[\Big]{\psi_3'}{v \, \iota_-(f(t))\cdot I[\psi_1\otimes\psi_2]}
	= \inner[\Big]{\psi_3'}{I\sqbrac[\big]{\copropts{w}{v\, f(t)} \cdot (\psi_1\otimes \psi_2)}}. \label{eqn:jac_complexanalytic_sigma}
\end{align}
In particular, $I$ intertwines the action of $v_n = v\, t^n$ on $M_3$
and that of $\copropts{w}{v_n}$ on $M_1\otimes M_2$.
The case for elements such as $v\, f(t)$, with $f(t)=t^a(t-w)^b$ and $b<0$, is more subtle and hence more interesting.
While \eqref{eqn:jac_complexanalytic_sigma} continues to hold,
we may also expand the term $\iota_- \left(f(t)\right)$ on the left-hand side to get
\begin{align}
	\inner[\Big]{\psi_3'}{I\sqbrac[\big]{\copropts{w}{v\, f(t)} \cdot (\psi_1\otimes \psi_2)}}
	&= \inner[\Big]{\psi_3'}{v \, \iota_-(t^a (t-w)^b)\cdot I[\psi_1\otimes\psi_2]} \nonumber\\
	&= \sum_{m=0}^{\infty} \binom{b}{m} (-w)^m \inner[\Big]{\psi_3'}{v\, t^{a+b-m}\cdot I[\psi_1\otimes\psi_2]}\nonumber\\
	&= \sum_{m=0}^{\infty} \binom{b}{m} (-w)^m \inner[\Big]{\psi_3'}{I\sqbrac[\big]{\copropts{w}{v\, t^{a+b-m}} \cdot (\psi_1\otimes \psi_2)}}\nonumber\\
	&= \inner[\Big]{\psi_3'}{I\sqbrac[\big]{\copropts{w}{v\, \iota_-(f(t))} \cdot (\psi_1\otimes \psi_2)}},
\end{align}
assuming that everything converges.  If so, we can summarise this simply as saying that the subspace generated by
\begin{align}
\copropts{w}{v\, f(t)} \cdot (\psi_1\otimes \psi_2) - \copropts{w}{v\, \iota_-(f(t))} \cdot (\psi_1\otimes \psi_2)
\end{align}
is in the kernel of any $P(w)$-intertwining map $I$.

This has a most natural interpretation: If we act on elements $\psi_1 \otimes \psi_2$ in the $P(w)$-tensor product with $v f(t)$ or with its expansion about $t=\infty$, then the result must be the same:
\begin{equation} \label{eq:consistency}
	\copropts{w}{v \, f(t)} = \copropts{w}{v \, \iota_-(f(t))} \qquad \text{on} \qquad M_1\fus{P(w)}M_2.
\end{equation}
The natural guess for a model of the universal $P(w)$-tensor product of $M_1$ and $M_2$ is therefore
\begin{align} \label{eq:P(w)def}
	M_1\fus{P(w)}M_2 = \dfrac{M_1\otimes M_2}{\ang[\Big]{\brac[\Big]{\copropts{w}{v \, f(t)} - \copropts{w}{v \, \iota_-(f(t))}}(M_1 \otimes M_2)}},
\end{align}
where the quotient is by
the vector space spanned obtained by taking
all $v\in V$ and $f(t)\in R_{P(w)}$.
Moreover, there is only one natural way to turn this space into a $V$-module, namely by letting $v_n$ act as $\copropts{w}{v \, t^n}$.  Notice that this quotient requires us to have extended the action of the modes of $V$ on $M_1 \otimes M_2$ to an action of the ``global mode algebra'' $V \otimes R_{P(w)}$.  This globalisation avoids the need to explicitly translate the action of the modes between different local coordinates, as in \eqref{eq:setcoprodequal}.

Comparing the definition \eqref{eq:P(w)def} with Gaberdiel's definition \eqref{eq:Gabdef} for the fusion product $M_1 \fus{} M_2$, we find striking similarities as well as one major difference.  The latter is the fact that \eqref{eq:Gabdef} involves two distinct coproduct actions, deduced from locality, while here there is only one.  To explain this, recall that \eqref{eq:P(w)def} was derived from the Cauchy-Jacobi identity \eqref{eqn:IntwMapJacobi}, the intertwining map analogue of the Jacobi identity of vertex algebras, and that the Jacobi identity naturally subsumes both locality and the \ope{}.  It is therefore natural to expect that we can recover from this formalism the identities that were derived using locality in \cref{sec:NGK}.  Indeed, we shall confirm these expectations shortly.

First, however, let us provide some concrete formulae that capture the $\coprosymb_w$-action and
quotient relations of \eqref{eq:P(w)def}.  This will show that this definition of $M_1\fus{P(w)}M_2$ essentially coincides with Gaberdiel's definition of the fusion product $M_1 \fus{} M_2$. In the next \lcnamecref{sec:doubledual}, we shall provide dual versions of these formulae.  From the action \eqref{eqn:coprod} with $f(t) = t^n$, we obtain
\begin{align}
	v_n (\psi_1\otimes \psi_2)
	&= \copropts{w}{v\, t^n} \cdot (\psi_1\otimes \psi_2)
	= \brac[\big]{v \, \iota_+ T_w(t^n)\cdot \psi_1} \otimes \psi_2 + \psi_1\otimes \brac[\big]{v \, \iota_+(t^n) \cdot \psi_2} \notag \\
	&= \brac[\big]{v \, \iota_+((t+w)^n)\cdot \psi_1} \otimes \psi_2 + \psi_1\otimes \brac[\big]{v \, t^n \cdot \psi_2} \notag \\
	&= \sum_{m=0}^{\infty} \binom{n}{m} w^{n-m} (v_m \psi_1) \otimes \psi_2 + \psi_1\otimes (v_n \psi_2).
\end{align}
This is identical to Gaberdiel's action, given in \eqref{eq:ngkcoproduct+ve} and \eqref{eq:ngkspeccoproduct2-ve}, with $w_1 = w$ and $w_2 = 0$.\footnote{We recall that the action of a $P(w)$-intertwining map on $\psi_1 \otimes \psi_2$ is (a projection of) $\psi_1(w) \psi_2$, in physics notation, explaining this specialisation of insertion points.} The actions of $V$ on $M_1 \fus{} M_2$ and $M_1\fus{P(w)}M_2$ therefore coincide.
As before, the sum acting on $\psi_1$ is actually finite due to the definition of a module over a \voa{}.

If we explicitly compute the action \eqref{eqn:coprod} with $f(t) = (t-w)^n$, we instead arrive at Gaberdiel's other action, given in \eqref{eq:ngkcoproduct+ve} and \eqref{eq:ngkspeccoproduct1-ve}, with $w_1 = 0$ and $w_2 = -w$:
\begin{equation}
	\copropts{w}{v\, (t-w)^n} \cdot (\psi_1\otimes \psi_2)
	= (v_n \psi_1) \otimes \psi_2 + \sum_{m=0}^{\infty} \binom{n}{m} (-w)^{n-m} \psi_1 \otimes (v_m \psi_2).
\end{equation}
Again, this is very reasonable as $t^n$ and $(t-w)^n$ are related by a rigid translation by $w$ while Gaberdiel's two coproducts are also identified up to a rigid translation by $w$, see \eqref{eq:setcoprodequal} (which assumes that $w=1$).  The icing on the cake is the fact that the relation
\begin{equation} \label{eq:icing}
	\copropts{w}{v \, (t-w)^n} = \copropts{w}{v \, \iota_-((t-w)^n)} = \sum_{m=0}^{\infty} \binom{n}{m} (-w)^m \copropts{w}{v_{n-m}},
\end{equation}
imposed by the definition \eqref{eq:P(w)def} of $M_1\fus{P(w)}M_2$, is now seen to reduce to Gaberdiel's translation identity \eqref{eq:coprodidentity} (once $w$ is set to $1$).  The latter is of course equivalent to the relations that are imposed by his definition \eqref{eq:Gabdef} of $M_1 \fus{} M_2$.

It should now be clear that the (formal) manipulations of this \lcnamecref{sec:models} amount to a second derivation of Gaberdiel's definition \eqref{eq:Gabdef} of the fusion product, here called the $P(w)$-tensor product.  Unfortunately, this means that the fruits of this labour suffer from exactly the same problems as before:
\begin{enumerate}
	\item The most urgent problem is that
	\begin{equation}
		\copropts{w}{v \, \iota_- (t^a (t-w)^b)} = \sum_{m=0}^{\infty} \binom{b}{m} (-w)^m \copropts{w}{v_{a+b-m}}
	\end{equation}
	involves an infinite sum of coproducts if $b$ is negative. As noted in \eqref{eq:doesnotconverge}, substituting in the coproduct actions on the \rhs{} gives hopelessly divergent results.
	\item A second problem is that we have not restricted the targets $W_3$ of our intertwining maps to lie in the category $\categ{C}$ that is provided to us.  This means that we may have fewer intertwining operators when a category is specified, thereby implying that the tensor product so-defined may be smaller than it might otherwise be.  In particular, the model \eqref{eq:P(w)def} for $M_1 \fus{P(w)} M_2$ need not lie in $\categ{C}$.
	As mentioned above, we shall ignore such categorical considerations in this note in order to focus on the algorithmic aspects.
	\item More fundamentally, we have not yet addressed the question of whether this model is actually a $V$-module.  If so, then it still remains to construct a universal intertwining map $\fus{P(w)}$ in order to complete the identification with the universal definition of \cref{sec:mathdef}.
\end{enumerate}
Nevertheless, we now have the advantage of having rephrased Gaberdiel's definition of the fusion product in a language that is mathematically more precise: that of intertwining maps.  We shall exploit this in the following \lcnamecref{sec:doubledual} when we turn to	the rigorous ``double dual approach'' of HLZ \cite{HLZ}.  This is perhaps the cleanest way to get around the first difficulty mentioned above and thus facilitate addressing the remaining problems.  We mention that Nahm \cite{NahQua94} was also well aware of this utility of dual spaces.

\section{The Huang-Lepowsky-Zhang approach: Double duals} \label{sec:doubledual}

In \eqref{eq:defvsdual}, we defined the restricted dual $M'$ of a graded $V$-module $M$ ($V$ being a \voa{} as usual) as a vector space, promising that we would in time equip $M'$ with the structure of a $V$-module.  That time has now come and the way to equip $M'$ is through an involutive antiautomorphism ${}^\opp$ on the modes $v_n$, $n \in \ZZ$, of $V$ that acts as a kind of adjoint for the canonical pairing of $M'$ and $M$:
\begin{align}
	\inner*{v_n \psi'}{\psi} = \inner*{\psi'}{v_n^{\opp} \psi} \qquad \text{($\psi \in M$, $\psi' \in M'$).}
	\label{eqn:dualreq}
\end{align}
This turns $M'$ into a $V$-module.

It is shown in \cite{FHL} that a natural choice for this involution, for completely general \voas{}, is given by the following formula.
Using mathematics notation for modes and assuming that $v$ has conformal weight $h_v$, the formula is
\begin{subequations} \label{eq:defopp}
	\begin{align}
		(v_n^{(\mathfrak{m})})^{\opp} = (-1)^{h_v}\sum_{j\ge 0} \dfrac{1}{j!}(L(1)^jv)_{-j+2h_v -2-n}^{(\mathfrak{m})}
		\label{eqn:dualmath}.
	\end{align}
	The formula looks much nicer with physicists' notation:
	\begin{align}
		(v_n^{(\mathfrak{p})})^{\opp} = (-1)^{h_v}\sum_{j\ge 0} \dfrac{1}{j!}(L(1)^jv)_{-n}^{(\mathfrak{p})}
		=(-1)^{h_v} (\ee^{L(1)}v)_{-n}^{(\mathfrak{p})}.
	\end{align}
\end{subequations}
We now illustrate this formula with examples in order to clarify this definition for physicists.

\begin{example}
	Suppose that $v$ is \emph{quasiprimary}, meaning in particular that it is annihilated by $L(1)$.  Then,
	\begin{equation} \label{eq:oppquasi}
		(v_n^{(\mathfrak{p})})^{\opp} = (-1)^{h_v} v_{-n}^{(\mathfrak{p})}.
	\end{equation}
	If we extend ${}^{\opp}$ to the field $v(z) = \sum_n v_n^{(\mathfrak{p})} z^{-n-h_v}$, then we have
	\begin{equation}
		v(z)^{\opp} = (-1)^{h_v} z^{-2h_v} v(z^{-1}).
	\end{equation}
	Apart from the sign, which we shall discuss in the following examples, this adjoint for the quasiprimary field $v(z)$ will be very familiar to physicists.
\end{example}

\begin{example}
	Interesting special cases of \eqref{eq:oppquasi} include the following:
	\begin{itemize}
		\item The identity operator $\wun = \vacuum_0^{(\mathfrak{p})}$ is self-adjoint, meaning that $\wun^{\opp} = \wun$, as expected.
		\item The Virasoro modes $L(n) = \confvec_n^{(\mathfrak{p})}$ satisfy $L(n)^{\opp} = L(-n)$, as expected.  In particular, $L(0)$ is self-adjoint.
		\item The modes of a weight-$1$ quasiprimary field, for example an affine current of the $\SLG{SU}{2}_k$ \wzw{} model, satisfy $J_n^{(\mathfrak{p})} = -J_{-n}^{(\mathfrak{p})}$, $J=E,H,F$.  In particular, the Cartan zero mode $H_0^{(\mathfrak{p})}$ is \emph{not} self-adjoint and the adjoint of $E_n^{(\mathfrak{p})}$ is \emph{not} $F_{-n}^{(\mathfrak{p})}$, contrary (perhaps) to expectations.
	\end{itemize}
	This possibly unexpected behaviour for affine modes is easily explained by noting that one can always twist the definition of ${}^{\opp}$ by composing with an automorphism of $V$.  In the case where $V = \SLG{SU}{2}_k$, we can twist by the finite Weyl reflection of $\SLA{sl}{2}$, which acts as $E \mapsto -F$, $H \mapsto -H$ and $F \mapsto -E$, to recover the expected adjoint.\footnote{In terms of the classification of real simple Lie algebras using involutions, the definition \eqref{eq:defopp} corresponds to the split real form, while the adjoint familiar to physicists is associated with the compact real form.}
\end{example}

\begin{example}
	The case where $V$ is the Heisenberg (free boson) \voa{} and $v = a_{-1}^{(\mathfrak{p})} \vacuum$ is the generator likewise gives $(a_0^{(\mathfrak{p})})^{\opp} = -a_0^{(\mathfrak{p})}$, which can also be ``fixed'' by twisting by the automorphism $a \mapsto -a$ of $\SLA{gl}{1}$.  However, we can appreciate the utility of this sign, and thus of the definition \eqref{eq:defopp}, by considering (as in the Coulomb gas formalism) the modified conformal vector
	\begin{equation}
		\confvec_{\lambda} = \frac{1}{2} (a_{-1}^{(\mathfrak{p})})^2 \vacuum + \frac{1}{2} \lambda a_{-2}^{(\mathfrak{p})} \vacuum \qquad \text{($\lambda \in \RR$)}
	\end{equation}
	and its associated Virasoro zero modes $L_{\lambda}(0)$.  For $\lambda = 0$, a straightforward computation shows that $(a_n^{(\mathfrak{p})})^{\opp} = -a_{-n}^{(\mathfrak{p})}$ and $(a_n^{(\mathfrak{p})})^{\opp} = +a_{-n}^{(\mathfrak{p})}$ both give $L_0(n)^{\opp} = L_0(-n)$, as required.  However, if $\lambda \neq 0$, then $v$ is no longer quasiprimary and \eqref{eq:defopp} instead gives
	\begin{equation} \label{eq:itjustworks}
		(a_n^{(\mathfrak{p})})^{\opp} = \lambda \delta_{n,0} \wun - a_{-n}^{(\mathfrak{p})}.
	\end{equation}
	Another straightforward calculation now shows that the general Ansatz $(a_n^{(\mathfrak{p})})^{\opp} = \alpha \delta_{n,0} \wun + \beta a_{-n}^{(\mathfrak{p})}$ is only consistent with $L_{\lambda}(n)^{\opp} = L_{\lambda}(-n)$ if $\alpha = \lambda$ and $\beta = -1$, as in \eqref{eq:itjustworks}.
\end{example}

We now wish to enlarge this $\opp$ adjoint
to be defined on the bigger space $V\otimes R_{P(w)}$.
In fact, we will go even further and define it on $V\otimes\CC[[t,t^{-1}]]$.
Inserting $v_n = v \, t^n$ into \eqref{eqn:dualmath}, so switching back to mathematician's conventions, we obtain
\begin{equation}
	(v \, t^n)^{\opp} = \sum_{m=0}^{\infty} \frac{t^{-j} L(1)^j}{j!} (-t^2)^{h_v} v \, t^{-2-n} = \ee^{t^{-1} L(1)} (-t^2)^{L(0)} v \, t^{-2-n}.
\end{equation}
We shall therefore define
\begin{equation}
	v^{\opp}=\ee^{t^{-1}L(1)}(-t^2)^{L(0)} v\,t^{-2}
\end{equation}
and then extend $\opp$ linearly to $V\otimes\CC[[t,t^{-1}]]$ by
\begin{equation}
	(v\, f(t))^{\opp} = v^{\opp} f(t^{-1}).
\end{equation}
It is not very hard to prove that $\brac*{(v\, f(t))^{\opp}}^{\opp} = v\, f(t)$, hence that ${}^{\opp}$ is still involutive.

\begin{example}
	Before moving on, let us rewrite the affine and Virasoro examples discussed above in this new notation.  For affine modes $J_n = J \, t^n$, we have
	\begin{align}
		J^{\opp} = e^{t^{-1}L(1)}(-t^2)^{L(0)}Jt^{-2} = -J,
	\end{align}
	hence $J_n^{\opp} = (J \, t^n)^{\opp} = J^{\opp} t^{-n} = -J \, t^{-n} = -J_{-n}$, as before.  For Virasoro modes $L(n) = \confvec \, t^{n+1}$, we have instead
	\begin{align}
		\confvec^{\opp} &= e^{t^{-1}L(1)}(-t^2)^{L(0)}\confvec t^{-2} = \confvec \, t^2
	\end{align}
	and so $L(n)^{\opp} = (\confvec \, t^{n+1})^{\opp} = \confvec^{\opp} t^{-n-1} = \confvec \, t^{-n+1} = L(-n)$, also as before.
\end{example}

We are now ready to transfer the $\coprow{w}$-action from the previous \lcnamecref{sec:models} to dual modules.  Consider the full dual $(M_1\otimes M_2)^\ast$. While this space is undoubtedly too big, we recall that the notion dual to taking a quotient of $M_1\otimes M_2$ is identifying an appropriate subspace of $(M_1\otimes M_2)^\ast$.  This subspace shall be identified after we have dualised the action
\eqref{eqn:coprod} of $V \otimes R_{P(w)}$, remembering to replace $t$ by $t^{-1}$, and have thus obtained
an action of $V\otimes\CC[t^{-1},t,(t^{-1}-w)^{-1}]$ on $(M_1\otimes M_2)^{\ast}$.  Note that because $(t^{-1} - w)^{-1} = -w^{-1} \, t (t-w^{-1})^{-1}$, we may regard this as an action of $V \otimes R_{P(w^{-1})}$.

Let $\psi^\ast$ be a generic element of $(M_1\otimes M_2)^\ast$. For $v \in V$ and $f(t) \in R_{P(w^{-1})}$, we define
\begin{multline}
	\inner[\big]{v\, f(t)\cdot \psi^\ast}{\psi_1\otimes \psi_2} =
	\inner[\big]{\psi^\ast}{\copropts{w}{(v\, f(t))^{\opp}} \cdot (\psi_1\otimes \psi_2)} \\
	=\inner*{\psi^\ast}{\brac*{v^{\opp}\, \iota_+ T_w \brac[\big]{f(t^{-1})}\cdot \psi_1}\otimes \psi_2}
	+ \inner*{\psi^\ast}{\psi_1\otimes \brac*{v^{\opp}\, \iota_+ \brac[\big]{f(t^{-1})}\cdot \psi_2}},
\end{multline}
for all $\psi_1 \in M_1$ and $\psi_2 \in M_2$.  Again, we pause to give examples.

\begin{example}
	In the case of affine modes, this dual action of $V \otimes R_{P(w^{-1})}$ specialises to
	\begin{align}
		\inner*{J_n\cdot \psi^\ast}{\psi_1\otimes \psi_2}
		&=\inner*{J\, t^n\cdot \psi^\ast}{\psi_1\otimes \psi_2} \notag \\
		&=\inner[\Big]{\psi^*}{\brac[\big]{-J\, \iota_+T_w (t^{-n})\cdot \psi_1} \otimes \psi_2}
		+ \inner[\Big]{\psi^*}{\psi_1\otimes \brac[\big]{-J\, \iota_+(t^{-n})\cdot \psi_2}} \notag \\
		&= -\inner[\Big]{\psi^*}{\brac[\big]{J\, \iota_+((t+w)^{-n})\cdot \psi_1} \otimes \psi_2}
		- \inner[\Big]{\psi^*}{\psi_1\otimes \brac[\big]{J\, t^{-n}\cdot \psi_2}} \notag \\
		&= -\sum_{m=0}^{\infty} \binom{-n}{m} w^{-n-m} \inner*{\psi^*}{(J_m \psi_1) \otimes \psi_2} - \inner*{\psi^*}{\psi_1 \otimes (J_{-n} \psi_2)}.
	\end{align}
	The Virasoro version is
	\begin{multline}
		\inner*{L(n)\cdot \psi^\ast}{\psi_1\otimes \psi_2}
		=\inner*{\confvec\, t^{n+1}\cdot \psi^\ast}{\psi_1\otimes \psi_2} \\
		= \sum_{m=0}^{\infty} \binom{-n+1}{m} w^{-n+1-m} \inner*{\psi^*}{(L(m-1) \psi_1) \otimes \psi_2} + \inner*{\psi^*}{\psi_1 \otimes (L(-n) \psi_2)}.
	\end{multline}
	A few special cases are worth noting:
	\begin{subequations}
		\begin{align}
			\inner{L(1)\cdot\psi^\ast}{\psi_1\otimes \psi_2 }
			&= \inner{\psi^\ast}{(L(-1) \psi_1)\otimes \psi_2 }
			+ \inner{\psi^\ast}{\psi_1\otimes (L(-1) \psi_2) }, \label{eq:L1HLZ} \\
			\inner{L(0)\cdot\psi^\ast}{\psi_1\otimes \psi_2 }
			&= w\,\inner{\psi^\ast}{(L(-1) \psi_1)\otimes \psi_2 }
			+ \inner{\psi^\ast}{(L(0) \psi_1)\otimes \psi_2 }
			+ \inner{\psi^\ast}{\psi_1\otimes (L(0) \psi_2) }, \label{eq:L0HLZ} \\
			\inner{L(-1)\cdot\psi^\ast}{\psi_1\otimes \psi_2 }
			&= w^2\,\inner{\psi^\ast}{(L(-1) \psi_1)\otimes \psi_2 }
			+2w\,\inner{\psi^\ast}{(L(0) \psi_1)\otimes \psi_2 } \notag \\
			&\mspace{100mu} + \inner{\psi^\ast}{(L(1) \psi_1)\otimes \psi_2 }
			+ \inner{\psi^\ast}{\psi_1\otimes (L(1) \psi_2) }. \label{eq:L-1HLZ}
		\end{align}
	\end{subequations}
\end{example}

Recall now from \eqref{eq:P(w)def} that we should impose the consistency relation
\begin{equation}
	\copropts{w}{v\, f(t)} \cdot \left(\psi_1\otimes  \psi_2\right) = \copropts{w}{v\, \iota_{-}(f(t))} \cdot \left(\psi_1\otimes \psi_2\right) \qquad \text{($f(t)\in R_{P(w)}$).}
\end{equation}
More precisely, we should impose its dual version, namely that we should restrict to the subspace of those $\psi^\ast \in (M_1 \otimes M_2)^*$ for which
\begin{align}
	\inner*{\psi^\ast}{\copropts{w}{v\, f(t)} \cdot \left(\psi_1\otimes  \psi_2\right)}
	= \inner*{\psi^\ast}{\copropts{w}{v\, \iota_{-}(f(t))}\cdot\left(\psi_1\otimes \psi_2\right)},
\end{align}
for all $v \, f(t) \in R_{P(w)}$, $\psi_1 \in M_1$ and $\psi_2 \in M_2$.  Equivalently, the desired subspace consists of those $\psi^\ast$ for which
\begin{align}
	v^{\opp}\, f(t^{-1}) \cdot \psi^\ast = v^{\opp}\, \iota_+(f(t^{-1})) \cdot \psi^\ast,
\end{align}
for all $v \, f(t) \in R_{P(w)}$, because expanding about $t=\infty$ ($\iota_-$) is dual to expanding about $t=0$ ($\iota_+$).  Since ${}^{\opp}$ is involutive, this consistency condition may be written in the form
\begin{subequations} \label{eq:P(w)compatibility}
	\begin{equation} \label{eqn:P(w)compat}
		v \, f(t) \cdot \psi^\ast = v \, \iota_+(f(t)) \cdot \psi^\ast \qquad \text{($v \, f(t) \in V \otimes R_{P(w^{-1})}$).}
	\end{equation}
	In other words, the dual action should be compatible with any infinite sums that are obtained by expanding $f(t)$ using $\iota_+$.  This is one of the \emph{$P(w)$-compatibility conditions} of HLZ, see \cite[Eq.~(5.141)]{HLZ}.  The other condition is in fact nothing more than requiring that
	\begin{align}
		v_n \cdot \psi^* = v\, t^n\cdot \psi^\ast = 0 \qquad \text{($v \in V$ and $n\gg0$).}
		\label{eqn:lowertrunc}
	\end{align}
\end{subequations}
This is, in general, required for the subspace of $(M_1 \otimes M_2)^*$ satisfying \eqref{eqn:P(w)compat} to stand any chance of being a $V$-module. Moreover, \eqref{eqn:lowertrunc} also guarantees that any infinite sum in \eqref{eqn:P(w)compat} is actually finite.  The condition \eqref{eqn:P(w)compat} is thus an equality in $(M_1 \otimes M_2)^*$, rather than in its completion.

The functionals that we are really after here are those that act as $\inner*{\psi^*}{\psi_1\otimes \psi_2} = \inner{\psi_3'}{I[\psi_1\otimes \psi_2]}$, where $I$ is any intertwining operator of type $\intwtype{M_3}{M_1}{M_2}$ ($M_3$ is some test module) and $\psi_3'$ is an arbitrary element of the restricted dual $M_3'$. These functionals are capable of acting non-trivially on large subspaces of $M_1\otimes M_2$. Therefore, the full dual $(M_1\otimes M_2)^\ast$ first provides us with enough room to work with such functionals (without having to deal with completions!) and second, the convergence issues that plagued our previous fusion product definitions, \eqref{eq:Gabdef} and \eqref{eq:P(w)def}, are neatly sidestepped by the natural ``lower truncation'' requirement \eqref{eqn:lowertrunc}.

Note that \eqref{eqn:P(w)compat} is dual to \eqref{eq:consistency} which, as we saw in \eqref{eq:icing}, is equivalent to Gaberdiel's ``coproduct equality'' \eqref{eq:setcoprodequal}.  In other words, the cure for the divergences that rendered Gaberdiel's definition of fusion meaningless is just \eqref{eqn:lowertrunc}. Note also that, as defined,
one must check the $P(w)$-compatibility condition \eqref{eqn:P(w)compat} for
all $v\, f(t) \in V \otimes R_{P(w^{-1})}$.
We can however do better!  Zhang essentially proved in \cite{ZhaKL08} that it is enough to check this condition for $v \in S$, where $S$ is a set of strong generators of $V$.

The HLZ definition for the $P(w)$-tensor product of the $V$-modules $M_1$ and $M_2$ can now be stated as follows:
\begin{itemize}
	\item Determine the subspace of $(M_1 \otimes M_2)^*$ consisting of elements $\psi^*$ that satisfy both $P(w)$-compatibility conditions \eqref{eq:P(w)compatibility}.
	\item In this subspace, consider the subspace of ``finite-energy'' vectors that are spanned by the generalised $L(0)$-eigenvectors.
	\item Define the $P(w)$-tensor product $M_1 \fus{P(w)} M_2$ to be the restricted dual of this finite-energy subspace.
\end{itemize}
This ``double dual'' approach then provides a candidate model for the universal $P(w)$-tensor product of \cref{def:univP(w)tp}.  There is one caveat: we have again ignored the category $\categ{C}$ completely.  In \cite{HLZ}, HLZ impose additional conditions beyond \eqref{eq:P(w)compatibility} so that the resulting $P(w)$-tensor product indeed lies in $\categ{C}$.  Because our aim is to compare with the physicists' fusion product, which is category-agnostic,
we do not discuss these details.

So far, we have discussed how this double dual approach overcomes or avoids the first two problems with the proposed model \eqref{eq:P(w)def}.  It therefore remains to prove that the space we have identified above as the $P(w)$-tensor product is actually a $V$-module.  This essentially boils down to questions about the given action of $V$ on $(M_1\otimes M_2)^\ast$ and the subspace
of compatible functionals.  Is this actually an action of $V$?  Is the subspace indeed stable under this action?

As noted in \cref{sec:NGK}, a proof that the action respects commutation rules and associativity was indicated by Gaberdiel for the case in which $V$ is a universal affine Kac-Moody algebra or the Virasoro algebra.  A completely general proof
is given by HLZ in \cite{HLZ}, lifting commutation rules to Borcherds identities (also known as generalised commutation relations), with no restrictions on the \voa{} or its module category.  There, it is also proved that the space of functionals $\psi^\ast$ satisfying the compatibility conditions \eqref{eq:P(w)compatibility} is indeed stable under the given action of $V \otimes R_{P(w^{-1})}$.  These proofs actually use formal variables and formal delta functions.  It would be interesting to prove them completely in the complex-analytic setting.

\medskip

We round out this section by going back to the example, discussed at the start of \cref{sec:models}, of tensor products for modules $M_1$ and $M_2$ over a commutative associative unital algebra $A$. We may regard $A$ as a vertex operator algebra with $Y(a,z)b=a\cdot b$ and $\confvec = 0$. In effect, $a$ is identified with the constant term $a_{-1} = a \, t^{-1}$ of $Y(a,z)$, all other terms being zero, and we have a trivial conformal structure: the central charge is $0$ and every vector has conformal weight $0$.  In addition, any $A$-module $M$ is naturally a module for this \voa{} via $Y_M(a,z)\psi = a\cdot \psi$.

What does the double dual construction of HLZ\footnote{The alert reader will notice that this double dual construction is overkill here because the fields $Y(a,z)$ are independent of $z$.  Nevertheless, we feel it helps to unpack this abstract machinery in the simplest case and see that it works.  We shall consider a less straightforward example in the next \lcnamecref{sec:virex}.} give for this class of \voas{}?  We first note that $(a\, f(t))^{\opp}=a\, t^{-2}f(t^{-1})$, so in particular we have $a^{\opp} = (a\,t^{-1})^{\opp} = a_{-1}\,t^{-1} = a$.  The $A$-action on the dual module $M^*$ is therefore defined by $\inner*{a\cdot \psi'}{\psi} = \inner*{\psi'}{a\cdot \psi}$, consistent with expectations.  The $\coprow{w}$-action on $M_1 \otimes M_2$ is therefore given by
\begin{equation}
	\copropts{w}{a\,t^n} \cdot (\psi_1\otimes \psi_2)
	= \brac[\big]{a\,\iota_+ T_w(t^n) \cdot \psi_1} \otimes \psi_2 + \psi_1 \otimes \brac[\big]{a\,\iota_+(t^n)} \cdot \psi_2
	= \delta_{n=-1} \, \psi_1 \otimes (a\cdot \psi_2),
\end{equation}
since the coefficient of $t^{-1}$ in the expansion of $(t+w)^n$ about $t=0$ is always $0$.  It follows that
\begin{equation} \label{eq:easierA-modcalc}
	\inner[\big]{a\, t^n \cdot \psi^\ast}{\psi_1\otimes \psi_2}
	= \inner[\big]{\psi^\ast}{\copropts{w}{a\, t^{-n-2}} \cdot (\psi_1\otimes \psi_2)}
	= \delta_{n=-1} \, \inner[\big]{\psi^\ast}{\psi_1 \otimes (a\cdot \psi_2)}.
\end{equation}
A similar, but more involved computation, gives
\begin{multline}
	\inner[\big]{a\, t^j(t^{-1}-w)^k \cdot \psi^\ast}{\psi_1\otimes \psi_2}
	=
	\binom{-j-2}{-k-1}w^{k-j-1} \inner*{\psi^\ast}{(a\cdot \psi_1) \otimes \psi_2} \\
	+
	\binom{k}{j+1} (-w)^{k-j-1} \inner*{\psi^\ast}{\psi_1 \otimes (a \cdot \psi_2)},
\end{multline}
where we note the commonly employed convention that $\binom{n}{r} = 0$ if $r$ is a negative integer.

However, the $P(w)$-compatibility condition \eqref{eqn:P(w)compat} lets us take a different route by first using $\iota_+$ to expand $t^j(t^{-1}-w)^k$ on the \lhs{} and then using \eqref{eq:easierA-modcalc}:
\begin{align}
	\inner[\big]{a\, t^j(t^{-1}-w)^k \cdot \psi^\ast}{\psi_1\otimes \psi_2}
	&= \sum_{m=0}^{\infty} \binom{k}{m} (-w)^m \inner[\big]{a \, t^{j-k+m} \cdot \psi^\ast}{\psi_1\otimes \psi_2} \notag \\
	&=
	\binom{k}{k-j-1} (-w)^{k-j-1} \inner[\big]{\psi^\ast}{\psi_1\otimes (a\cdot \psi_2)}.
\end{align}
Therefore, we must have
\begin{align} \label{eq:theeqnforA-mods}
	\binom{-j-2}{-k-1} (-1)^{k-j-1} \inner*{\psi^\ast}{(a\cdot \psi_1) \otimes \psi_2}
	= \sqbrac*{\binom{k}{k-j-1} - \binom{k}{j+1}} \inner[\big]{\psi^\ast}{\psi_1\otimes (a\cdot \psi_2)}.
\end{align}
Note first that the \lhs{} is $0$ if $k \ge 0$, while the binomial coefficients on the \rhs{} are either equal (if $0 \le j+1 \le k$) or both $0$ (for all other $j$).  We may therefore restrict to $k<0$ and consider the following four possibilities for $j$:
\begin{enumerate}
	\item $j \le k-1$ and $j \ge -1$ is impossible, so at most one of the binomial coefficients on the \rhs{} may be non-zero.
	\item $j > k-1$ and $j < -1$ is possible, but then $-k-1 > -j-2 \ge 0$ and both sides are zero.
	\item If $j \le k-1$ and $j < -1$, then the second binomial coefficient on the \rhs{} is zero.  Since $-j-2 \ge -k-1 \ge 0$, we have
	\begin{equation}
		\binom{-j-2}{-k-1} (-1)^{k-j-1} = \binom{-j-2}{k-j-1} (-1)^{k-j-1} = \binom{k}{k-j-1}
	\end{equation}
	and so \eqref{eq:theeqnforA-mods} reduces to
	\begin{equation}
		\inner*{\psi^\ast}{(a\cdot \psi_1) \otimes  \psi_2}=\inner*{\psi^\ast}{\psi_1\otimes (a\cdot \psi_2)}.
	\end{equation}
	\item A similar calculation for $j > k-1$ and $j \ge -1$ likewise reduces to this same equation.
\end{enumerate}
It is quite enlightening to redo this analysis using generating functions (thereby considering all values of $j$ and $k$ simultaneously) and the properties of formal delta functions, as in \cite{HuaLeLiZha06}.  Either way, we conclude that the $P(w)$-compatibility conditions for commutative associative unital algebras pick out precisely the $\psi^*$ that implement the constraint that reduces $M_1 \otimes M_2$ to $M_1 \otimes_A M_2$.  The double dual model of the HLZ $P(w)$-tensor product is, for these \voas{}, just the usual tensor product over $A$.
There are other quirks about this example, mainly arising from the fact that the central charge is zero, which are discussed in detail in \cite{HuaLeLiZha06}.

\section{An explicit example of a fusion calculation: Virasoro at $c=-2$}
\label{sec:virex}

Let $V$ be the universal vertex operator algebra associated with the Virasoro algebra at central charge $c=-2$. At this central charge, the universal Virasoro vertex operator algebra is actually simple.  Let $M$ be the irreducible \hw{} Virasoro module of central charge $-2$ whose \hwv{} $\gamma$ has conformal weight $-\frac{1}{8}$.  It follows \cite{FreVer92} that $M$ is a $V$-module.  Let $\psi_1$ and $\psi_2$ be arbitrary elements of $M$ and, finally, let $\psi^{\ast}$ be an arbitrary element of $(M\otimes M)^*$.

Our aim is, obviously, to calculate the fusion product $M \fus{} M$, from here on identified with the $P(w)$-tensor product $M \fus{P(w)} M$ (with $w \in \CC^{\times}$ arbitrary).  We shall do so using the formalism of HLZ, identifying $M \fus{} M$ as a Virasoro module.  For this, we need to determine the possibilities for the $\psi^\ast$ that satisfy HLZ's $P(w)$-compatibility
conditions \eqref{eq:P(w)compatibility}.  In fact, we will only derive some restrictions on $\psi^\ast$ that follow as consequences of
being $P(w)$-compatible. It is in general a hard task to prove that such $\psi^\ast$ indeed exist.

Experts will note that this example has historical significance, being the very first calculation performed \cite{GabInd96} using the computational method that is now known as the \ngk{} algorithm.  The motivation behind this first calculation was, of course, the expectation that the fusion product would exhibit a non-diagonalisable action of $L(0)$, a fact that had been previously been established by Gurarie \cite{GurLog93} using correlation functions.  We shall discuss the calculation of \cite{GabInd96}, and this algorithm, in \cref{sec:algngk}.  In what follows, one may take $w=1$ for convenience, but this is not at all necessary. We will also freely use the notations that were introduced in previous sections.

Consider $v\, f(t)\in V\otimes R_{P(w^{-1})}$.  By a result of Zhang \cite{ZhaKL08}, it is enough for us to take $\xi = \confvec\, f(t)$, since $\confvec$ strongly generates $V$.  By way of preparation, we compute the actions
\begin{align} \label{eq:vir1}
	\inner*{L(n)\cdot \psi^\ast}{\psi_1\otimes\psi_2}
	&= \inner*{\psi^*}{\copropts{w}{(\confvec\, t^{n+1})^{\opp}} \cdot (\psi_1\otimes \psi_2)}
	= \inner*{\psi^*}{\copropts{w}{\confvec\, t^{-n+1}} \cdot (\psi_1\otimes \psi_2)} \notag \\
	&= \inner*{\psi^*}{\brac*{\confvec\, \iota^+ \brac[\big]{(t+w)^{-n+1}} \cdot \psi_1}\otimes \psi_2}
		+ \inner*{\psi^*}{\psi_1 \otimes \brac*{\confvec\, \iota^+(t^{-n+1}) \cdot \psi_2}} \notag \\
	&= \sum_{m=0}^{\infty} \binom{-n+1}{m} w^{-n-m+1} \inner[\big]{\psi^\ast}{(L(m-1) \psi_1) \otimes \psi_2}
		+ \inner[\big]{\psi^\ast}{\psi_1\otimes (L(-n) \psi_2)}
\end{align}
and (in the same way)
\begin{multline} \label{eq:vir2}
	\inner*{\confvec\, t(t^{-1}-w)^n \cdot \psi^\ast}{\psi_1\otimes \psi_2} \\
	= \inner*{\psi^\ast}{\brac*{\brac[\Big]{w L(n-1) + L(n)} \psi_1} \otimes \psi_2}
		+ \sum_{m=0}^{\infty} \binom{n}{m} (-w)^{n-m} \inner*{\psi^\ast}{\psi_1\otimes (L(m) \psi_2)}.
\end{multline}

To begin with, let us assume that $\psi^\ast \in (M \otimes M)^*$
satisfies both the $P(w)$-compatibility conditions \eqref{eq:P(w)compatibility} and, in addition,
$L(n)\cdot\psi^\ast =0$ for all $n\ge 1$.
\begin{enumerate}
	\item For $n\ge 1$, \eqref{eq:vir1} gives
	\begin{align}
		0 &= \inner*{L(n)\cdot \psi^\ast}{\psi_1\otimes\psi_2} \notag \\
		&= \sum_{m=0}^{\infty} \binom{-n+1}{m} w^{-n-m+1} \inner[\big]{\psi^\ast}{(L(m-1) \psi_1) \otimes \psi_2}
			+ \inner[\big]{\psi^\ast}{\psi_1\otimes (L(-n) \psi_2)}.
	\end{align}
	This implies that the values $\inner*{\psi^\ast}{M\otimes \psi_2}$ are determined by the values $\inner*{\psi^\ast}{M\otimes \gamma}$, because the action of any negative mode $L(-n)$ in the second tensor factor may be traded for an action on the first tensor factor.  (We recall that $\gamma$ is the highest weight vector of $M$.)
	\item Now we investigate $\confvec\, t (t^{-1}-w)^n \cdot \psi^\ast$ for $n\le -1$. On the one hand, the $P(w)$-compatibility condition \eqref{eqn:P(w)compat} requires that
	\begin{align}
		\confvec \,t (t^{-1}-w)^n \cdot \psi^\ast
		= \confvec \,\iota_+ \brac[\big]{t (t^{-1}-w)^n} \cdot \psi^\ast
		= \sum_{m=0}^{\infty} \binom{n}{m} (-w)^m L(m-n) \cdot \psi^\ast
		=0,
	\end{align}
	because $m-n$ is strictly positive. On the other hand, combining this with \eqref{eq:vir2} now gives
	\begin{align}
		w \inner[\Big]{\psi^\ast}{\brac[\big]{L(n-1) \psi_1} \otimes \gamma}
			+ \inner[\Big]{\psi^\ast}{\brac[\big]{L(n) \psi_1} \otimes \gamma}
			+ (-w)^n \inner[\Big]{\psi^\ast}{\psi_1\otimes \brac[\big]{L(0) \gamma}} = 0.
	\label{eqn:usefulforL0}
	\end{align}
	Since we have assumed that $n\le -1$, the action of $L(-2), L(-3), \ldots$ on $\psi_1$ may be traded for that of $L(-1), L(-2), \ldots$ respectively.  It therefore follows
	that $\inner*{\psi^\ast}{M\otimes \gamma}$ is determined by the numbers $\inner[\big]{\psi^\ast}{(L(-1)^k \gamma) \otimes \gamma}$, for $k \in \ZZ_{\ge 0}$.  However, the \hwv{} $\gamma$ satisfies
	\begin{equation} \label{eq:c=-2SV}
		L(-1)^2 \gamma = \frac{1}{2} L(-2) \gamma
	\end{equation}
	by virtue of $M$ being irreducible \cite{AstStr97}.  Thus, $\inner*{\psi^\ast}{M\otimes \gamma}$ is actually determined by the numbers $\inner*{\psi^\ast}{\gamma\otimes \gamma}$ and $\inner*{\psi^\ast}{(L({-1})\gamma)\otimes \gamma}$.
	Consequently, the space of $P(w)$-compatible $\psi^\ast$ satisfying $L(n) \cdot \psi^\ast=0$, for all $n\ge 1$, is at most $2$-dimensional.
	\item We now investigate the action of $L(0)$ on such $\psi^\ast$. As $L(n) L(0) \cdot \psi^\ast=0$, for all $n\ge 1$, and the $P(w)$-compatible elements form a $V$-submodule \cite{HLZ},
	we can apply all the discussion above with $L(0) \cdot \psi^\ast$ in place of $\psi^\ast$.  In particular, \eqref{eq:L0HLZ} gives
	\begin{subequations}
		\begin{align} \label{eq:comparelater}
			\inner*{L(0) \cdot \psi^\ast}{\gamma\otimes \gamma}
			&= \inner[\big]{\psi^\ast}{(wL({-1})\gamma)\otimes \gamma + (L(0)\gamma)\otimes \gamma + \gamma\otimes (L(0)\gamma)} \notag \\
			&=-\frac{1}{4}\inner*{\psi^\ast}{\gamma\otimes \gamma}+w\inner*{\psi^\ast}{(L(-1)\gamma)\otimes \gamma}
		\intertext{and, using also \eqref{eq:c=-2SV}, then \eqref{eqn:usefulforL0} with $n=-1$ and $\psi_1=\gamma$,}
			\inner*{L(0)\cdot\psi^\ast}{(L({-1})\gamma)\otimes \gamma}
			&= \inner[\big]{\psi^\ast}{(wL({-1})^2\gamma)\otimes \gamma + (L(0)L(-1)\gamma)\otimes \gamma) + (L(-1)\gamma)\otimes (L(0)\gamma)} \notag \\
			&= \frac{1}{2}w\inner*{\psi^\ast}{(L(-2)\gamma)\otimes \gamma} + \frac{3}{4}\inner*{\psi^\ast}{(L(-1)\gamma)\otimes \gamma} \notag \\
			&= \frac{1}{4}\inner*{\psi^\ast}{(L(-1)\gamma)\otimes \gamma} + \frac{1}{2} w^{-1} \inner*{\psi^\ast}{\gamma\otimes(L(0)\gamma)} \notag \\
			&= -\frac{1}{16} w^{-1} \inner*{\psi^\ast}{\gamma\otimes\gamma} + \frac{1}{4}\inner*{\psi^\ast}{(L(-1)\gamma)\otimes \gamma}.
		\end{align}
	\end{subequations}
	\item In effect, we have shown that
	\begin{align} \label{eq:matrix}
		\begin{bmatrix}
			\inner*{L(0)\cdot\psi^\ast}{\gamma\otimes \gamma}\\
			\inner*{L(0)\cdot\psi^\ast}{(L(-1)\gamma)\otimes \gamma}
		\end{bmatrix}
		=
		\begin{bmatrix}
			-1/4 & w \\
			-w^{-1}/16 & 1/4
		\end{bmatrix}
		\begin{bmatrix}
			\inner*{\psi^\ast}{\gamma\otimes \gamma}\\
			\inner*{\psi^\ast}{(L(-1)\gamma)\otimes \gamma}
		\end{bmatrix}
		.
	\end{align}
	The $2\times2$ matrix representing this action of $L(0)$ has zero trace and zero determinant, hence consists of a rank $2$ Jordan block with eigenvalue $0$.  However, recall that we have not shown that there are two linearly independent $P(w)$-compatible $\psi^*$ that are annihilated by the $L(n)$ with $n\ge1$, just that there are at most $2$.  Consequently, this space of $\psi^*$ could actually be $1$- or even $0$-dimensional.

	In principle, we should test these $\psi^*$ for $P(w)$-compatibility using $\confvec \, f(t)$, for completely general $f(t) \in R_{P(w^{-1})}$.  However, it is very difficult to show that no further constraints arise.  Instead, one can resort to information obtained through indirect means.  In particular, Gurarie showed in \cite{GurLog93} that the correlation function involving four copies of the field $\gamma(z) = Y_{\fus{}}(\gamma,z)$ (at four different insertion points) possesses logarithmic singularities.  As this cannot happen if $L(0)$ acts diagonalisably on the fusion product $M \fus{} M$, this strongly suggests that there are no further constraints.  We shall assume the truth of this statement, hence that we have correctly identified a rank $2$ Jordan block in the $L(0)$-action.  It is useful to indicate this conclusion pictorially as follows:
	\begin{align}
	\begin{tikzpicture}[->,>=latex,scale=1.5,baseline={(l.base)}]
		\node (l) at (0,0) {$\bullet$};
		\node (r) at (1,0) {$\bullet$};
		\draw (r) -- node[above] {$\scriptstyle L(0)$} (l);
	\end{tikzpicture}
	.
	\end{align}
	We emphasise that this picture indicates just a part of the space $F$ of of all functionals on $M \otimes M$ that satisfy the $P(w)$-compatibility conditions \eqref{eq:P(w)compatibility}.  \cref{eq:matrix} demonstrates that this part describes finite-energy vectors in $F$.
	\item We next consider $L(-1) \cdot \psi^\ast$, assuming that $\psi^*$ is $P(w)$-compatible and that $L(n) \cdot \psi^\ast=0$ for all $n\ge0$.  In other words, we wish to compute the action of $L(-1)$ on the $\psi^*$ that corresponds to the $L(0)$-eigenvector in the Jordan block identified above.  However, the $P(w)$-compatible elements form a $V$-module \cite{HLZ} and
	\begin{equation}
		L(n)L(-1) \cdot \psi^\ast = L(-1)L(n) \cdot \psi^\ast + (n+1) L(n-1) \cdot \psi^*=0,
	\end{equation}
	for all $n\ge 1$.
	The above discussion therefore applies \emph{mutatis mutandis} with $L(-1) \cdot \psi^\ast$ in place of $\psi^\ast$.
	In particular, the conclusion that such $\psi^*$ are generalised $L(0)$-eigenvectors of eigenvalue $0$ must apply to $L(-1) \cdot \psi^\ast$.  As this is impossible (the $L(0)$-eigenvalue is clearly $1$), the only way out is to have  $\inner*{L({-1})\psi^\ast}{M\otimes M}$ are determined by the numbers
	$L(-1) \cdot \psi^\ast=0$.

	This conclusion does not apply if we relax the condition that $L(0) \cdot \psi^* = 0$, that is if we consider $\psi^*$ to be one of the generalised eigenvectors of $L(0)$, because $L(1) L(-1) \cdot \psi^* = 2L(0) \cdot \psi^* \neq 0$.  To identify if $L(-1) \cdot \psi^*$ is zero or not, we would need to identify a basis of the subspace of $P(w)$-compatible elements that are annihilated by the $L(n)$ with $n\ge2$.  This can be done, though we shall not do so here, and the result is that $L(-1) \cdot \psi^*$ is not zero when $L(0) \cdot \psi^* \neq 0$.  We add this information to our pictorial representation of $F$ thusly:
	\begin{align} \label{diag:virex}
		\begin{tikzpicture}[->,>=latex,scale=1.5,baseline={(mid.base)}]
			\node (mid) at (0,-0.5) {\vphantom{M}};
			\node (a) at (0,0) {$\bullet$};
			\node (b) at (2,0) {$\bullet$};
			\node (c) at (0,-1) {$\times$};
			\node (d) at (2,-1) {$\bullet$};
			\draw (b) -- node[above] {$\scriptstyle L(0)$} (a);
			\draw[dashed] (a) -- node[left] {$\scriptstyle L(-1)$} (c);
			\draw (b) -- node[right] {$\scriptstyle L(-1)$} (d);
			\draw (d) -- node[below left] {$\scriptstyle \frac{1}{2} L(1)$} (a);
		\end{tikzpicture}
		\quad.
	\end{align}
	Here, the $\times$ indicates that the target state is $0$.
	\item Using the well-known structure theory for Virasoro \hwms{}, we conclude that the submodule of $F$ generated by the $L(0)$-eigenvector of eigenvalue $0$ is irreducible and is therefore isomorphic to $M(0) / M(1)$, where $M(h)$ denotes the $c=-2$ Virasoro Verma module of conformal weight $h$.  In particular, it is isomorphic to the vacuum module $V$ of the \voa{}.  The generalised $L(0)$-eigenvectors do not generate \hwms{} of course.  However, each becomes a genuine eigenvector in the quotient module $F/V$ and therefore generates a \hwm{} $H$ of conformal weight $0$ in this quotient.  Unfortunately, the information we have does not allow us to identify $H$ --- all we can say is that it is not isomorphic to $V$.

	To identify $H$, we must delve deeper into the structure of the fusion product.  Considering the Virasoro action on the subspace of $P(w)$-compatible $\psi^*$ that are annihilated by the $L(n)$ with $n\ge4$, some tedious computation verifies that there is a second vanishing relation akin to the relation $L(-1) \cdot \psi^* = 0$ established above.  The image of this relation in $F/V$ is
	\begin{equation}
		\brac*{L(-1)^2 - 2 L(-2)} L(-1) \cdot [\psi^*] = 0,
	\end{equation}
	where $[\psi^*]$ is the image of any generalised $L(0)$-eigenvector of conformal weight $0$.  This vanishing, along with the structure theory of Virasoro \hwms{} now identifies $H$ as the quotient $M(0) / M(3)$.

	It is therefore natural to conjecture that the finite-energy submodule $\underline{F}$ of $F$ is characterised by the following non-split short exact sequence of $V$-modules:
	\begin{equation}
		\dses{\frac{M(0)}{M(1)}}{\underline{F}}{\frac{M(0)}{M(3)}}.
	\end{equation}
	Taking restricted duals gives the following non-split short exact sequence for the fusion product:
	\begin{equation}
		\dses{\brac*{\frac{M(0)}{M(3)}}'}{M \fus{} M}{\frac{M(0)}{M(1)}}.
	\end{equation}
	Here, we have noted that $M(0)/M(1) \cong V$ is irreducible, hence self-dual.  This can, however, be reorganised so as to arrive at the same non-split short exact sequence as for $\underline{F}$:
	\begin{equation} \label{es:MxM}
		\dses{\frac{M(0)}{M(1)}}{M \fus{} M}{\frac{M(0)}{M(3)}}.
	\end{equation}
	We indicate this reorganisation pictorially by taking the dual (reversing the arrows) in \eqref{diag:virex}:
	\begin{equation}
		\begin{tikzpicture}[->,>=latex,scale=1.5,baseline={(mid.base)}]
			\node (mid) at (0,-0.5) {\vphantom{M}};
			\node (a) at (0,0) {$\bullet$};
			\node (b) at (2,0) {$\bullet$};
			\node (d) at (2,-1) {$\bullet$};
			\draw (a) -- node[above] {$\scriptstyle L(0)$} (b);
			\draw (d) -- node[right] {$\scriptstyle L(1)$} (b);
			\draw (a) -- node[below left] {$\scriptstyle \frac{1}{2} L(-1)$} (d);
			\begin{scope}[shift={(3.15,0)}]
				\node at (0,-1/2) {$\equiv$};
			\end{scope}
			\begin{scope}[shift={(4,0)}]
				\node (a) at (0,0) {$\bullet$};
				\node (b) at (2,0) {$\bullet$};
				\node (d) at (2,-1) {$\bullet$};
				\draw (b) -- node[above] {$\scriptstyle L(0)$} (a);
				\draw (b) -- node[right] {$\scriptstyle \frac{1}{2} L(-1)$} (d);
				\draw (d) -- node[below left] {$\scriptstyle L(1)$} (a);
			\end{scope}
		\end{tikzpicture}
		\quad.
	\end{equation}

	If this conjecture is true, then the sequence \eqref{es:MxM}, along with the non-diagonalisable $L(0)$-action, makes $M \fus{} M$ a \emph{staggered module} \cite{RohRed96} and the structure theory of such modules \cite{RidSta09} shows that this non-split sequence completely determines the (self-dual) fusion product $M \fus{} M$ up to isomorphism (the corresponding extension group is $\CC$).  To verify this, one needs to rule out the possibility that the fusion product is in fact larger than this staggered module.  This can be done with the information currently available, but requires much deeper knowledge of extension groups (in a generalisation of category $\categ{O}$ that allows for non-diagonalisable $L(0)$-actions).
\end{enumerate}

With regard to this calculation, there are several points that merit further explanation.

First, note that at every stage, we are only giving \emph{necessary constraints} that are consequences of the
$P(w)$-compatibility conditions. This means that at every stage, it may be possible that we have fewer possibilities
for the $\psi^\ast$ than one might na\"{\i}vely expect.  Typically, such additional constraints arise from relations satisfied in the modules being fused, for instance the vanishing of singular vectors, and their interpretation using $P(w)$-compatibility.
In the physics literature, such extraneous $\psi^\ast$ are (dual versions of) the
\emph{spurious states} of Nahm \cite{NahQua94}, see \cref{sec:algngk}.

Second, recall that HLZ identify a subspace of the (enormous) full dual $(M_1\otimes M_2)^\ast$ as the restricted dual of the actual fusion product $M_1 \fus{} M_2$.  It seems like dualising this inclusion leads to the conclusion that the fusion product may be realised as a quotient of $M_1\otimes M_2$, vindicating the approaches discussed in \cref{sec:NGK,sec:models}.  However, we know that these approaches led to unacceptable divergences.  The crucial fact to pinpoint here is that ``dualising this inclusion'' is the source of our confusion --- the full dual $(M_1\otimes M_2)^\ast$ is not graded by generalised $L(0)$-eigenspaces, so we cannot take its restricted dual.  The best that we could conclude then is that the fusion product may be realised as the finite-energy vectors in the double dual $(M_1\otimes M_2)^{*\ast}$ (which is even more enormous than the full dual).

Lastly, in order to build fusion products algorithmically, for example by considering subspaces of $P(w)$-compatible $\psi^*$ that are annihilated by the $L(n)$ with $n \ge d$, for some $d$, we are naturally led to incorporate filtrations of the mode algebra into the procedure.  The obvious filtration is given by conformal weights, but we may have to get more creative in general.  We shall have a little more to say on this in the following \lcnamecref{sec:algngk}.

\section{The fusion algorithm of Nahm-Gaberdiel-Kausch} \label{sec:algngk}

We have reviewed Gaberdiel's original approach to defining fusion and shown that it may be interpreted mathematically in terms of $P(w)$-intertwining maps.  The resulting ``construction'' of the fusion product was shown to be formally meaningless due to convergence issues, but we have seen how this can be made rigorous using dual spaces \emph{\'{a} la} HLZ.  We have even seen how to perform non-trivial calculations in the HLZ formalism that allow one to identify fusion products, in favourable circumstances.  Now, we would like to complete the circle and discuss how physicists perform these calculations.  The method that we shall explain here was originally proposed by Nahm \cite{NahQua94} and was subsequently generalised and implemented by Gaberdiel and Kausch \cite{GabInd96}.  It is therefore known as the \ngk{} fusion algorithm, or NGK algorithm for short.

We shall discuss this algorithm shortly with the aid of the same example (Virasoro at $c=-2$) that was treated in the previous \lcnamecref{sec:virex}.  Here, we shall be fairly brief because there are already several detailed discussions addressing these practicalities in the physics literature, see for example \cite{GabInd96,EbeVir06,CreLog13,CanRasRidNS15}.  The point is really to make manifest that the NGK algorithm is essentially the dual of the computational method that we have outlined above for the HLZ formalism.

An elephantine question now enters the room: How did NGK develop a practical algorithm to construct fusion products when they knew that their definition of fusion suffered from divergences?  The answer is of course that physicists have a long history of dealing with divergences, especially when field theory is involved, and so their reaction to this seemingly intransigent block was decidedly meh.  Indeed, Nahm pointed out \cite{NahQua94} that these divergences could be fixed by working in the dual space.  More interestingly, however, he chose to ignore this in favour of a practical approach that avoided duals and instead worked with quotients.

To explain his idea, recall that our HLZ computation above began by considering the \emph{subspace} of $P(w)$-compatible $\psi^*$ in the dual of the tensor product space that were annihilated by the (dual) action of the $L(n)$ with $n\ge1$.  Dualising this now leads us to consider the \emph{quotient} of the tensor product space in which we impose annihilation by the action of the $L(-n) = L(n)^{\opp}$ with $n\ge1$.  On this quotient, we also need to impose the dual of $P(w)$-compatibility.  We have already noted above that the dual of \eqref{eqn:P(w)compat} is Gaberdiel's equality of coproducts \eqref{eq:setcoprodequal}, itself derived from locality, which is formally divergent on the full tensor product.  Let us check how this condition fares on Nahm's quotient.  Substituting \eqref{eq:ngkcoproduct+ve} or \eqref{eq:ngkcoproduct1-ve}, with $w_1 = 0$ and $w_2 = -1$, into \eqref{eq:coprodidentity} gives (for $v = \confvec$)
\begin{equation} \label{eq:gabdepth}
	(L(n) \otimes \wun) + \sum_{m=0}^{\infty} \binom{n+1}{m} (-1)^{n+1-m} (\wun \otimes L(m-1)) = \sum_{m=0}^{\infty} (-1)^m \binom{n+1}{m} \coproipts{2}{1,0}{L(n-m)}.
\end{equation}
The infinite sum on the \lhs{} is harmless when acting on any equivalence class $[\psi_1 \otimes \psi_2]$ because $v_m \psi_2 = 0$ for $m \gg 0$, by definition of a $V$-module.  The novel feature is thus that the infinite sum on the \rhs{} is now also rendered finite because we are working in a quotient in which negative modes act as $0$.  Nahm's quotient has thus cured the divergences in much the same way that \eqref{eqn:lowertrunc} did for HLZ.  The divergences are similarly cured for the other $v$ in this Virasoro \voa{} thanks to the magic of normal ordering.

Quotienting by the action of the $L(-n)$ with $n \ge 1$ is perfectly fine for identifying fusion products in rational \cft{}.  However, we saw in the previous \lcnamecref{sec:virex} that there are fusion products for which this is not going to be true.  Gaberdiel and Kausch recognised this in the course of studying such examples (which arise in so-called \emph{logarithmic} \cft{}) and therefore generalised Nahm's cure to incorporate quotients by the action of the $L(-n)$ with $n>d$, for some ``depth'' $d$.\footnote{In fact, they took this a step further and discussed quotients by actions of fairly arbitrary subalgebras of the mode algebra of $V$.  Appropriate filtrations by such subalgebras then lead to a consistent framework in which one can evaluate the action of any given mode.  The need for quite exotic filtrations is best exemplified by referring to the rather difficult computations that arise when studying fusion products for modules over non-rational affine \voas{}, see \cite{GabFus01,RidFus10}.}

We digress briefly to note that for a general \voa{} $V$, the depth-$0$ quotient of any irreducible $V$-module (whose conformal weights are bounded below) coincides with the image of said module under the Zhu functor of \cite{ZhuMod96}.  On the quotient, one can only compute the action of those modes of $V$ that commute with $L(0)$ (zero modes with physics conventions).  This action of course agrees with that of the Zhu algebra.  As regards depth-$d$ quotients, aficionados of higher-level Zhu algebras should feel right at home.  However, it is possible to define actions of certain non-zero modes that map between quotients of different depths \cite{GabInd96}.  This does not seem to have been incorporated into higher Zhu theory yet (though perhaps it should).

A natural question to ask now is whether one can honestly define fusion products in terms of these filtered quotients on which the divergence malady has been eradicated.  In favourable cases, such as that of the previous \lcnamecref{sec:virex} where one has the \hw{} theory of the Virasoro algebra at one's disposal, we only need analyse a finite number of these quotients in order to completely identify the fusion product.  However, an abstract definition should apply more generally.  Such a definition might go as follows:
\begin{itemize}
	\item Assemble the quotients into a projective system and take the projective limit.
	\item Define the fusion product to be the submodule of finite-energy vectors in this projective limit.
\end{itemize}
This definition is hinted at in Nahm's original paper \cite{NahQua94} and is proposed concretely in a paper of Tsuchiya and Wood \cite{TsuTen13} (though this has not been developed further to the best of our knowledge), see also similar work of Miyamoto \cite{MiyC114}.  Unfortunately, none of these papers seem to prove that this definition agrees with that of HLZ, even in favourable cases.  We are very tempted to conjecture that it does.

One obvious difficulty with this approach is in determining whether the result is independent of the choice of (suitable) filtration and its quotients.  For $C_2$-cofinite \voas{}, one might expect this to be the case with the filtration by conformal weight being perhaps sufficient to completely identify the results.  However, applying the NGK algorithm to non-rational affine \voas{} \cite{GabFus01,RidFus10} suggests strongly that this is a much more subtle question in general --- the presence of sectors that are twisted by spectral flow automorphisms means that filtering by conformal weight definitely does not suffice.  In fact, it can happen that all conformal weight filtration quotients lead to zero.

At the end of the day however, what is clear is that we may dualise the methodology we detailed in \cref{sec:virex} to compute fusion products in the HLZ formalism.  We shall explain in the rest of this \lcnamecref{sec:algngk} that this dual formalism is essentially the NGK algorithm.  The conclusion is that fusion computations performed using NGK will necessarily agree with those performed using HLZ, despite the fact that the NGK formalism currently has no rigorous definition for the fusion product.  Of course, the NGK algorithm has the relative advantage of dispensing with the abstraction of the dual space formalism.

As Feynman advised, we shall now shut up and calculate, again considering the $c=-2$ Virasoro fusion product of $M$ with itself, as discussed in the previous \lcnamecref{sec:virex}.  We begin by investigating the depth-$0$ quotient of $M \otimes M$, imposing relations such as Gaberdiel's coproduct equality \eqref{eq:gabdepth} (which we recall is dual to $P(w)$-compatibility) in order to cut it down to the depth-$0$ quotient of $M \fus{} M$.  If we take $n\le-1$, then the \rhs{} acts as $0$ on the depth-$0$ quotient and we have
\begin{equation} \label{eq:1<->2}
	(L(n) \otimes \wun) = \sum_{m=0}^{\infty} \binom{n+1}{m} (-1)^{n-m} (\wun \otimes L(m-1)).
\end{equation}
For $n\le-2$, this lets us swap the action of $L(n)$ on the first tensor factor for a linear combination of terms in which $L(-1)$, $L(0)$, $L(1)$ and so on act on the second tensor factor.  By swapping all such actions onto the second factor, we are left with only $L(-1)$-modes acting on the first.  However, two of these may be replaced by an $L(-2)$, by \eqref{eq:c=-2SV}, which may then be swapped for an action on the second factor.  It follows that (the images of) $\gamma \otimes M$ and $(L(-1) \gamma) \otimes M$ together form a spanning set for the zero-depth quotient.

We can do even better by using \eqref{eq:ngkspeccoproduct2-ve} for the action of the negative Virasoro modes, again because this action is required to vanish.  This formula simplifies, for $n\le-1$, to
\begin{equation} \label{eq:2<->1}
	(\wun \otimes L(n)) = -\sum_{m=0}^{\infty} \binom{n+1}{m} (L(m-1) \otimes \wun),
\end{equation}
which allows us to swap the action of a negative mode of the second factor for $L(-1)$, $L(0)$, $L(1)$ and so on acting on the first.  With some careful bookkeeping, it is easy to see that by bouncing the actions from one factor to the other and back again, we get the spanning set
\begin{equation} \label{eq:spanset}
	\set{\gamma \otimes \gamma, (L(-1) \gamma) \otimes \gamma}
\end{equation}
for the zero-depth quotient of $M \fus{} M$.

As in \cref{sec:virex}, it is not easy to tell if this spanning set is a basis or not, but we can rephrase our conclusion in a more general (and hopefully enlightening) manner.  Let us call the span of $\gamma$ and $L(-1) \gamma$ the \emph{special subspace} $M^{\text{ss}}$ of $M$.  We shall denote the depth-$0$ quotient of an $L(0)$-graded module $N$ by $N^{(0)}$.  The conclusion is then that the depth-$0$ quotient of $M \fus{} M$ may be expressed as a quotient of the tensor product of the special subspace of $M$ and its depth-$0$ quotient:
\begin{equation}
	M^{\text{ss}} \otimes M^{(0)} \lsra (M \fus{} M)^{(0)}.
\end{equation}
As these are just vector spaces, it would also be reasonable to consider the depth-$0$ quotient of the fusion product as a subspace of the tensor product.  Nahm's first deep insight into fusion was to realise that this is a quite general fact.  For an arbitrary \voa{} $V$ and an arbitrary $L(0)$-graded $V$-module $M$, let $C_1(M)$ denote the image in $M$ of the action of the negative modes $v_n$, for all $v \in V$ and $n\ge1$ (using mathematics conventions!).  The \emph{special subspace} $M^{\text{ss}}$ is then defined \cite{NahQua94} to be the vector space quotient $M / C_1(M)$.  Nahm then argues that
\begin{equation} \label{eq:NahmFusQuot}
	M^{\text{ss}} \otimes N^{(0)} \lsra (M \fus{} N)^{(0)},
\end{equation}
for $V$-modules $M$ and $N$.

In the example at hand, we can determine the action of $L(0)$ on the candidate space for $(M \fus{} M)^{(0)}$ by applying \eqref{eq:ngkcoproduct+ve} and bouncing the actions back and forth between the tensor factors until we again have a linear combination of the elements of \eqref{eq:spanset}.  For example, \eqref{eq:ngkcoproduct+ve}, \eqref{eq:c=-2SV}, \eqref{eq:1<->2} and \eqref{eq:2<->1} give
\begin{align}
	\copropts{1,0}{L(0)} (L(-1) \gamma \otimes \gamma)
	&= (L(-1)^2 \gamma \otimes \gamma) + (L(0) L(-1) \gamma \otimes \gamma) + (L(-1) \gamma \otimes L(0) \gamma) \notag\\
	&= \frac{1}{2} (L(-2) \gamma \otimes \gamma) + \frac{3}{4} (L(-1) \gamma \otimes \gamma) \notag \\
	&= \frac{1}{2} \sqbrac[\Big]{(\gamma \otimes L(-1) \gamma) + (\gamma \otimes L(0) \gamma)} + \frac{3}{4} (L(-1) \gamma \otimes \gamma) \notag \\
	&= -\frac{1}{16} (\gamma \otimes \gamma) + \frac{1}{4} (\gamma \otimes L(-1) \gamma),
\end{align}
which of course agrees with the result obtained in \eqref{eq:comparelater} using the HLZ double dual formalism.  Here, as there, we find a non-diagonalisable action of $L(0)$ which, when combined with Gurarie's observation about logarithmic singularities in correlation functions, shows that \eqref{eq:spanset} is a basis for $(M \fus{} M)^{(0)}$.

This game of bouncing actions between tensor factors is not confined to the depth-$0$ world.  Gaberdiel and Kausch showed in \cite{GabInd96} that \eqref{eq:NahmFusQuot} generalises readily to depth-$d$, at least for the Virasoro case with filtration by conformal weight:
\begin{equation}
	M^{\text{ss}} \otimes N^{(d)} \lsra (M \fus{} N)^{(d)}.
\end{equation}
This has since been verified for several other \voas{} \cite{GabRat96,GabFus97,GabFus01,RidBos14,CanRasRidNS15}.  As in \cref{sec:virex}, we shall look quickly at what happens to our example when $d=1$.  First, note that $\set{\gamma, L(-1) \gamma}$ is a basis for both $M^{\text{ss}}$ and $M^{(1)}$, hence that $(M \fus{} M)^{(1)}$ is a quotient of a four-dimensional space.  We saw in \cref{sec:virex} that this fusion quotient was actually only three-dimensional.  There, this was established using some straightforward abstract reasoning.  Here, we want to take the time to see how it also follows from a relation satisfied by the Virasoro action on $M$.

Recall that to identify the special subspace of $M$, we used the \sv{} relation \eqref{eq:c=-2SV}.  This relation also holds for the second factor of the tensor product $M \otimes M$, hence may be exploited to deduce additional relations to impose on $M^{\text{ss}} \otimes M^{(1)}$.  In particular, $\copropts{1,0}{L(-1)}^2 = 0$ and \eqref{eq:c=-2SV} give
\begin{align} \label{eq:spstate}
	0 &= \copropts{1,0}{L(-1)}^2 (\gamma \otimes \gamma)
	= (L(-1)^2 \gamma \otimes \gamma) + 2 (L(-1) \gamma \otimes L(-1) \gamma) + (\gamma \otimes L(-1)^2 \gamma) \notag \\
	&= \frac{1}{2} (L(-2) \gamma \otimes \gamma) + 2 (L(-1) \gamma \otimes L(-1) \gamma) + \frac{1}{2} (\gamma \otimes L(-2) \gamma).
\end{align}
The first term is simplified using \eqref{eq:gabdepth}:
\begin{equation}
	(L(-2) \gamma \otimes \gamma) = (\gamma \otimes L(-1) \gamma) + \frac{1}{8} (\gamma \otimes \gamma).
\end{equation}
The second is fine as is, so we simplify the third by using $\copropts{1,0}{L(-2)} = 0$:
\begin{equation}
	0 = \copropts{1,0}{L(-2)} (\gamma \otimes \gamma)
	= (L(-1) \gamma \otimes \gamma) + \frac{1}{8} (\gamma \otimes \gamma) + (\gamma \otimes L(-2) \gamma).
\end{equation}
\cref{eq:spstate} therefore becomes
\begin{equation} \label{eq:spstate'}
	2 (L(-1) \gamma \otimes L(-1) \gamma) - \frac{1}{2} (L(-1) \gamma \otimes \gamma) + \frac{1}{2} (\gamma \otimes L(-1) \gamma) = 0,
\end{equation}
an additional relation that reduces the dimensionality of $(M \fus{} M)^{(1)}$ from $4$ down to $3$.  Nahm refers to the \lhs{} of \eqref{eq:spstate'} as a \emph{spurious state}.

The full identification of the fusion product $M \fus{} M$ in the NGK formalism now proceeds in a similar fashion to the HLZ computation discussed in \cref{sec:virex}.  We will not repeat the details here, but instead comment on certain mild differences.  As we have seen, the determination of spurious states seemed somewhat easier with the HLZ formalism.  This is because the depth-$0$ part of the HLZ algorithm captured states in the (dual) fusion product that were annihilated by the positive modes, thereby characterising all \svs{} at once.  On the other hand, we remarked that some high-powered extension group results were needed to conclude that the HLZ identification did not miss states in the fusion product not generated by \hwvs{}.  By contrast, this conclusion is easy with the NGK formalism because the depth-$0$ calculation captures all states that are not obtained by acting with a negative mode.  It would be very interesting to combine the two formalisms and see if a hybrid algorithm could efficiently take advantage of these observations.  We shall not do so here of course, leaving such speculations for future work.

\section{A (very brief) summary of other approaches to fusion}
\label{sec:otherapproaches}

If one is working with rational conformal field theories, for which the representation theory of the underlying \voa{} $V$ is semisimple with finitely many simple objects,
then the fusion product has the simple form \eqref{eq:fusingmodules}.  In mathematical language, we can rewrite this in the form
\begin{align} \label{eq:genfusrule}
	M_i \fus{P(w)} M_j \cong \bigoplus_{k \in S} \dim \sqbrac*{\mathcal{N}_{P(w)} \intwtype{M_k}{M_i}{M_j}} M_k,
\end{align}
where the $M_k$, with $k\in S$, enumerate the irreducible $V$-modules up to isomorphism and $\mathcal{N}_{P(w)}$ denotes the space of $P(w)$-intertwining maps of the indicated type.\footnote{We caution that despite this formula, an explicit construction of the fusion product using $P(w)$-compatibility conditions, among other things, is required in the work of Huang and Lepowsky \cite{HuaTheI95,HuaTheII95,HuaTheIII95} in order to build a braided tensor structure.}
Here, we are assuming that the fusion coefficients appearing in \eqref{eq:genfusrule} are all finite.
Now the problem of finding the fusion product is equivalent to computing the various fusion coefficients (these dimensions).
This problem can be solved, in principle, using the technology of Zhu algebras, as was stated by Frenkel and Zhu \cite{FreVer92}, with the proof given by Li \cite{LiThesis94,LiNuc98}. The formula is as follows:
\begin{align}
	\mathcal{N}_{P(w)} \intwtype{M_k}{M_i}{M_j} \cong \hom_{A(V)}(A(M_i) \otimes_{A(V)} M_j^\topspace, A(M_k)).
\end{align}
Here, $A(V)$ denotes the Zhu algebra of $V$, $A(M)$ denotes the image of Zhu's functor from $V$-modules to $A(V)$-bimodules, and $M^{\topspace}$ is the left-$A(V)$-submodule of $M$ spanned by the vectors of minimal conformal weight. The hom-space in this formula corresponds to $A(V)$-bimodule homomorphisms. The proof of this theorem is fairly technical, but uses essentially familiar ideas: showing that the action of an intertwining operator is fully determined by its action on ``small enough'' spaces, namely $A(M_1)\otimes M_2^\topspace$. The hard part is of course going back: to build an intertwining operator consistently, given only its action on such small spaces.

During the course of his proof of the Frenkel-Zhu bimodule theorem, Haisheng Li in \cite{LiThesis94,LiNuc98} presented another ``abstract'' construction of the fusion product philosophically similar to the one we have given above. The setting is also non-logarithmic. The fused module $M_1\fus{}M_2$ is spanned by ``modes'' $(\psi_1)_n(\psi_2)$, where $\psi_1$ and $\psi_2$ run through the respective modules and $n$ runs over the complex numbers. Here, $(\psi_1)_n$ is the mode corresponding to the ``universal intertwining operator''. Li thus builds an abstract vector space spanned by all the entities $(\psi_1)_n(\psi_2)$ and cuts it down by imposing relations satisfied by intertwining operators. To the best of our knowledge, the analogous construction has not been carried out in the logarithmic setting. However, a generalisation of the Frenkel-Zhu theorem to the logarithmic world exists \cite{HuaYanLog12}.  Unsurprisingly, it involves the higher Zhu algebras of \cite{DonVer98}.

Huang and Lepowsky's first series of papers on the theory of fusion products 
for rational conformal field theories was, at least in part, inspired by 
Kazhdan and Lusztig's series of papers \cite{KL1,KL2,KL3,KL4} that provided a 
tensor structure on the category of ordinary modules over an affine Lie algebra 
at levels $k$ satisfying $k+\dcox\not\in \QQ_{\ge 0}$. For a comparison of the 
Huang-Lepowsky and Kazhdan-Lusztig approaches for these cases, we refer to 
\cite{ZhaKL08}. Note that certain aspects of HLZ's opus \cite{HLZ} are 
decidedly harder than in \cite{KL1,KL2,KL3,KL4} --- it requires much more 
effort to prove in general that the fused object is actually a module for the 
\emph{\voa{}} as compared to a \emph{Lie algebra}. However, Kazhdan and Lusztig 
were also able to prove that their categories close under fusion, while Huang 
and Lepowsky actually build their tensor product theory by \emph{assuming} that 
fusion closes on a suitable category. Currently, the most general result 
confirming such closure under fusion is from \cite{HuaC209} where it is proved 
that categories of finite length modules over a $C_2$-cofinite (also known as 
\emph{lisse}) \voa{} are closed under fusion (see also \cite{HuaApp17}).  
Results pertaining to certain non-$C_2$-cofinite situations are also available, 
see, for example, \cite{CreBra18}.

One can find many other approaches to fusion in the literature, some of which we have already mentioned.  Tsuchiya and Wood have announced \cite{TsuTen13} a rigorous theory of fusion in terms of a projective limit of NGK quotients.  However, the proofs have not yet appeared.  Miyamoto has constructed \cite{MiyC114} a similar theory in which it is asserted that $C_1$-cofinite modules close under fusion.  This bears a strong resemblance to the main result in Nahm's original paper \cite{NahQua94} in which $C_1$-cofiniteness goes by the name of quasirationality (the reader may recall that this term was mentioned in the quote in the introduction).

We also want to mention an approach that has been developed by the statistical physics community, see \cite{PeaLog06,ReaAss07} for example, in which one computes ``fusion products'' for finite discretisations of the \cft{} and then takes the continuum scaling limit in order to recover information about the actual fusion products.  This approach is currently far from rigorous, but there is a concrete proposal \cite{GaiLat13} for this discretised fusion based on categories of modules over the Temperley-Lieb algebras and their generalisations.  One of the main issues here is the fascinating link between the Temperley-Lieb and Virasoro algebras as dictated by scaling limits \cite{GaiLog13}.  Comparisons between the discretised and conformal fusion results in the logarithmic case \cite{MorKac15} indicates that this link exhibits subtle structure that is still poorly understood.  Nevertheless, fusion gives us a powerful tools to better understand scaling limits.  A rigorous theory of scaling limits would certainly be a jewel in the crown for mathematical physics.

\flushleft

\begin{thebibliography}{10}

\bibitem{NahQua94}
W~Nahm.
\newblock Quasirational fusion products.
\newblock {\em Int. J. Mod. Phys.}, B8:3693--3702, 1994.
\newblock \oldpreprint{9402039}{hep-th}.

\bibitem{FeiCoh88}
B~Feigin and D~Fuchs.
\newblock Cohomology of some nilpotent subalgebras of the {Virasoro} and
  {Kac}-{Moody} {Lie} algebras.
\newblock {\em J. Geom. Phys.}, 5:209--235, 1988.

\bibitem{MooSeiPoly88}
G~Moore and N~Seiberg.
\newblock Polynomial equations for rational conformal field theories.
\newblock {\em Phys. Lett.}, B212:451--460, 1988.

\bibitem{MooSeiCFT89}
G~Moore and N~Seiberg.
\newblock Classical and quantum conformal field theory.
\newblock {\em Comm. Math. Phys.}, 123:177--254, 1989.

\bibitem{FreVer92}
I~Frenkel and Y~Zhu.
\newblock Vertex operator algebras associated to representations of affine and
  {Virasoro} algebras.
\newblock {\em Duke Math. J.}, 66:123--168, 1992.

\bibitem{LiThesis94}
H~Li.
\newblock {\em Representation theory and tensor product theory for vertex
  operator algebras}.
\newblock PhD thesis, Rutgers University, 1994.
\newblock \oldpreprint{9406211}{hep-th}.

\bibitem{LiNuc98}
H~Li.
\newblock An analogue of the {Hom} functor and a generalized nuclear democracy
  theorem.
\newblock {\em Duke Math. J.}, 93:73--114, 1998.
\newblock \oldpreprint{9706012}{q-alg}.

\bibitem{GabFus94}
M~Gaberdiel.
\newblock Fusion in conformal field theory as the tensor product of the
  symmetry algebra.
\newblock {\em Int. J. Mod. Phys.}, A9:4619--4636, 1994.
\newblock \oldpreprint{9307183}{hep-th}.

\bibitem{GabFus94b}
M~Gaberdiel.
\newblock Fusion rules of chiral algebras.
\newblock {\em Nucl. Phys.}, B417:130--150, 1994.
\newblock \oldpreprint{9309105}{hep-th}.

\bibitem{KazAff91}
D~Kazhdan and G~Lusztig.
\newblock Affine {Lie} algebras and quantum groups.
\newblock {\em Int. Math. Res. Not.}, 1991:21--29, 1991.

\bibitem{KL1}
D~Kazhdan and G~Lusztig.
\newblock Tensor structures arising from affine {Lie} algebras. {I}.
\newblock {\em J. Amer. Math. Soc.}, 6:905--947, 1993.

\bibitem{KL2}
D~Kazhdan and G~Lusztig.
\newblock Tensor structures arising from affine {Lie} algebras. {II}.
\newblock {\em J. Amer. Math. Soc.}, 6:949--1011, 1993.

\bibitem{KL3}
D~Kazhdan and G~Lusztig.
\newblock Tensor structures arising from affine {Lie} algebras. {III}.
\newblock {\em J. Amer. Math. Soc.}, 7:335--381, 1994.

\bibitem{KL4}
D~Kazhdan and G~Lusztig.
\newblock Tensor structures arising from affine {Lie} algebras. {IV}.
\newblock {\em J. Amer. Math. Soc.}, 7:383--453, 1994.

\bibitem{HL-vtc}
Y-Z Huang and J~Lepowsky.
\newblock Tensor products of modules for a vertex operator algebra and vertex
  tensor categories.
\newblock In {\em Lie theory and geometry}, volume 123 of {\em Progr. Math.},
  pages 349--383. Birkh{\"a}user, Boston, 1994.
\newblock \oldpreprint{9401119}{hep-th}.

\bibitem{HuaTheI95}
Y-Z Huang and J~Lepowsky.
\newblock A theory of tensor products for module categories for a vertex
  operator algebra, {I}.
\newblock {\em Selecta Math. New Ser.}, 1(4):699--756, 1995.
\newblock \oldpreprint{9309076}{hep-th}.

\bibitem{HuaTheII95}
Y-Z Huang and J~Lepowsky.
\newblock A theory of tensor products for module categories for a vertex
  operator algebra, {II}.
\newblock {\em Selecta Math. New Ser.}, 1(4):757--786, 1995.
\newblock \oldpreprint{9309159}{hep-th}.

\bibitem{HuaTheIII95}
Y-Z Huang and J~Lepowsky.
\newblock A theory of tensor products for module categories for a vertex
  operator algebra. {III}.
\newblock {\em J. Pure Appl. Algebra}, 100(1-3):141--171, 1995.
\newblock \oldpreprint{9505018}{hep-th}.

\bibitem{GabInd96}
M~Gaberdiel and H~Kausch.
\newblock Indecomposable fusion products.
\newblock {\em Nucl. Phys.}, B477:293--318, 1996.
\newblock \oldpreprint{9604026}{hep-th}.

\bibitem{HuaLepSurvey13}
Y-Z Huang and J~Lepowsky.
\newblock Tensor categories and the mathematics of rational and logarithmic
  conformal field theory.
\newblock {\em J. Phys.}, A46:494009, 2013.
\newblock \preprint{1304.7556}{hep-th}.

\bibitem{GabInt00}
M~Gaberdiel.
\newblock An introduction to conformal field theory.
\newblock {\em Rep. Prog. Phys.}, 63:607--667, 2000.
\newblock \oldpreprint{9910156}{hep-th}.

\bibitem{CreLog13}
T~Creutzig and D~Ridout.
\newblock Logarithmic conformal field theory: beyond an introduction.
\newblock {\em J. Phys.}, A46:494006, 2013.
\newblock \preprint{1303.0847}{hep-th}.

\bibitem{HLZ}
Y-Z Huang, J~Lepowsky, and L~Zhang.
\newblock Logarithmic tensor product theory {I}--{VIII}.
\newblock \preprint{1012.4193}{math.QA}, \preprint{1012.4196}{math.QA},
  \preprint{1012.4197}{math.QA}, \preprint{1012.4198}{math.QA},
  \preprint{1012.4199}{math.QA}, \preprint{1012.4202}{math.QA},
  \preprint{1110.1929}{math.QA}, \preprint{1110.1931}{math.QA}.

\bibitem{KirqAn02}
A~Kirillov Jr and V~Ostrik.
\newblock On a $q$-analogue of the {McKay} correspondence and the {ADE}
  classification of $\mathfrak{sl}_2$ conformal field theories.
\newblock {\em Adv. Math.}, 171:183--227, 2002.
\newblock \oldpreprint{0101219}{math.QA}.

\bibitem{HuaBra15}
Y-Z Huang, A~Kirillov Jr, and J~Lepowsky.
\newblock Braided tensor categories and extensions of vertex operator algebras.
\newblock {\em Comm. Math. Phys.}, 337:1143--1159, 2015.
\newblock \preprint{1406.3420}{math.QA}.

\bibitem{CreTen17}
T~Creutzig, S~Kanade, and R~McRae.
\newblock Tensor categories for vertex operator superalgebra extensions.
\newblock \preprint{1705.05017}{math.QA}.

\bibitem{GurLog93}
V~Gurarie.
\newblock Logarithmic operators in conformal field theory.
\newblock {\em Nucl. Phys.}, B410:535--549, 1993.
\newblock \oldpreprint{9303160}{hep-th}.

\bibitem{EbeVir06}
H~Eberle and M~Flohr.
\newblock {Virasoro} representations and fusion for general augmented minimal
  models.
\newblock {\em J. Phys.}, A39:15245--15286, 2006.
\newblock \oldpreprint{0604097}{hep-th}.

\bibitem{RidPer07}
P~Mathieu and D~Ridout.
\newblock From percolation to logarithmic conformal field theory.
\newblock {\em Phys. Lett.}, B657:120--129, 2007.
\newblock \preprint{0708.0802}{hep-th}.

\bibitem{RidLog07}
P~Mathieu and D~Ridout.
\newblock Logarithmic {$M \left( 2,p \right)$} minimal models, their
  logarithmic couplings, and duality.
\newblock {\em Nucl. Phys.}, B801:268--295, 2008.
\newblock \preprint{0711.3541}{hep-th}.

\bibitem{RidPer08}
D~Ridout.
\newblock On the percolation {BCFT} and the crossing probability of {Watts}.
\newblock {\em Nucl. Phys.}, B810:503--526, 2009.
\newblock \preprint{0808.3530}{hep-th}.

\bibitem{RasCla11}
J~Rasmussen.
\newblock Classification of {Kac} representations in the logarithmic minimal
  models {$LM \left( 1,p \right)$}.
\newblock {\em Nucl. Phys.}, B853:404--435, 2011.
\newblock \preprint{1012.5190}{hep-th}.

\bibitem{MorKac15}
A~Morin-Duchesne, J~Rasmussen, and D~Ridout.
\newblock Boundary algebras and {Kac} modules for logarithmic minimal models.
\newblock {\em Nucl. Phys.}, B899:677--769, 2015.
\newblock \preprint{1503.07584}{hep-th}.

\bibitem{CanRasRidNS15}
M~Canagasabey, J~Rasmussen, and D~Ridout.
\newblock Fusion rules for the logarithmic {$N=1$} superconformal minimal
  models {I}: The {Neveu}-{Schwarz} sector.
\newblock {\em J. Phys.}, A48:415402, 2015.
\newblock \preprint{1504.03155}{hep-th}.

\bibitem{CanRidRam16}
M~Canagasabey and D~Ridout.
\newblock Fusion rules for the logarithmic {$N=1$} superconformal minimal
  models {II}: Including the {Ramond} sector.
\newblock {\em Nucl. Phys.}, B905:132--187, 2016.
\newblock \preprint{1512.05837}{hep-th}.

\bibitem{GabRat96}
M~Gaberdiel and H~Kausch.
\newblock A rational logarithmic conformal field theory.
\newblock {\em Phys. Lett.}, B386:131--137, 1996.
\newblock \oldpreprint{9606050}{hep-th}.

\bibitem{GabFus09}
M~Gaberdiel, I~Runkel, and S~Wood.
\newblock Fusion rules and boundary conditions in the $c=0$ triplet model.
\newblock {\em J. Phys.}, A42:325403, 2009.
\newblock \preprint{0905.0916}{hep-th}.

\bibitem{WooFus10}
S~Wood.
\newblock Fusion rules of the {$W \left( p,q \right)$} triplet models.
\newblock {\em J. Phys.}, A43:045212, 2010.
\newblock \preprint{0907.4421}{hep-th}.

\bibitem{TsuTen13}
A~Tsuchiya and S~Wood.
\newblock The tensor structure on the representation category of the {$\mathcal
  W_p$} triplet algebra.
\newblock {\em J. Phys.}, A46:445203, 2013.
\newblock \preprint{1201.0419}{hep-th}.

\bibitem{GabFus01}
M~Gaberdiel.
\newblock Fusion rules and logarithmic representations of a {WZW} model at
  fractional level.
\newblock {\em Nucl. Phys.}, B618:407--436, 2001.
\newblock \oldpreprint{0105046}{hep-th}.

\bibitem{RidFus10}
D~Ridout.
\newblock Fusion in fractional level $\widehat{\mathfrak{sl}} \left( 2
  \right)$-theories with $k=-\tfrac{1}{2}$.
\newblock {\em Nucl. Phys.}, B848:216--250, 2011.
\newblock \preprint{1012.2905}{hep-th}.

\bibitem{CreRel11}
T~Creutzig and D~Ridout.
\newblock Relating the archetypes of logarithmic conformal field theory.
\newblock {\em Nucl. Phys.}, B872:348--391, 2013.
\newblock \preprint{1107.2135}{hep-th}.

\bibitem{RidBos14}
D~Ridout and S~Wood.
\newblock Bosonic ghosts at $c=2$ as a logarithmic {CFT}.
\newblock {\em Lett. Math. Phys.}, 105:279--307, 2015.
\newblock \preprint{1408.4185}{hep-th}.

\bibitem{FLM}
I~Frenkel, J~Lepowsky, and A~Meurman.
\newblock {\em Vertex Operator Algebras and the Monster}, volume 134 of {\em
  Pure and Applied Mathematics}.
\newblock Academic Press, Boston, 1988.

\bibitem{KacVer96}
V~Kac.
\newblock {\em Vertex Algebras for Beginners}, volume~10 of {\em University
  Lecture Series}.
\newblock American Mathematical Society, Providence, 1996.

\bibitem{FreVer01}
E~Frenkel and D~Ben-Zvi.
\newblock {\em Vertex Algebras and Algebraic Curves}, volume~88 of {\em
  Mathematical Surveys and Monographs}.
\newblock American Mathematical Society, Providence, 2001.

\bibitem{LeLi}
J~Lepowsky and H~Li.
\newblock {\em Introduction to Vertex Operator Algebras and their
  Representations}, volume 227 of {\em Progress in Mathematics}.
\newblock Birkh{\"a}user, Boston, 2004.

\bibitem{FeiAnn92}
B~Feigin, T~Nakanishi, and H~Ooguri.
\newblock The annihilating ideals of minimal models.
\newblock {\em Int. J. Mod. Phys.}, A7:217--238, 1992.

\bibitem{ZhuMod96}
Y~Zhu.
\newblock Modular invariance of characters of vertex operator algebras.
\newblock {\em J. Amer. Math. Soc.}, 9:237--302, 1996.

\bibitem{GabFus97}
M~Gaberdiel.
\newblock Fusion of twisted representations.
\newblock {\em Int. J. Mod. Phys.}, A12:5183--5208, 1997.
\newblock \oldpreprint{9607036}{hep-th}.

\bibitem{MooCla89}
G~Moore and N~Seiberg.
\newblock Classical and quantum conformal field theory.
\newblock {\em Comm. Math. Phys.}, 123:177--254, 1989.

\bibitem{MiyC114}
M~Miyamoto.
\newblock {$C_1$}-cofiniteness and fusion products for vertex operator
  algebras.
\newblock In {\em Conformal field theories and tensor categories}, Math. Lect.
  Peking Univ., pages 271--279. Springer, Heidelberg, 2014.
\newblock \preprint{1305.3008}{math.QA}.

\bibitem{AbeRat04}
T~Abe, G~Buhl, and C~Dong.
\newblock Rationality, regularity, and {$C_2$}-cofiniteness.
\newblock {\em Trans. Amer. Math. Soc.}, 356(8):3391--3402, 2004.

\bibitem{AdaLat09}
D~Adamovi\'{c} and A~Milas.
\newblock Lattice construction of logarithmic modules for certain vertex
  algebras.
\newblock {\em Selecta Math. New Ser.}, 15:535--561, 2009.
\newblock \preprint{0902.3417}{math.QA}.

\bibitem{RidVer14}
D~Ridout and S~Wood.
\newblock The {Verlinde} formula in logarithmic {CFT}.
\newblock {\em J. Phys. Conf. Ser.}, 597:012065, 2015.
\newblock \preprint{1409.0670}{hep-th}.

\bibitem{FHL}
I~Frenkel, Y-Z Huang, and J~Lepowsky.
\newblock On axiomatic approaches to vertex operator algebras and modules.
\newblock {\em Mem. Amer. Math. Soc.}, 104:viii+64, 1993.

\bibitem{MilWea02}
A~Milas.
\newblock Weak modules and logarithmic intertwining operators for vertex
  operator algebras.
\newblock In {\em Recent developments in infinite-dimensional {Lie} algebras
  and conformal field theory}, volume 297 of {\em Contemporary Mathematics},
  pages 201--225. American Mathematical Society, 2002.
\newblock \oldpreprint{0101167}{math.QA}.

\bibitem{HuaBook97}
Y-Z Huang.
\newblock {\em Two-dimensional conformal geometry and vertex operator
  algebras}, volume 148 of {\em Progress in Mathematics}.
\newblock Birkh{\"a}user, Boston, 1997.

\bibitem{DonLep93}
C~Dong and J~Lepowsky.
\newblock {\em Generalized vertex algebras and relative vertex operators},
  volume 112 of {\em Progress in Mathematics}.
\newblock Birkh\"{a}user Boston, Inc., Boston, MA, 1993.

\bibitem{CKLR}
T~Creutzig, S~Kanade, A~Linshaw, and D~Ridout.
\newblock {Schur}-{Weyl} duality for {Heisenberg} cosets.
\newblock {\em Transform. Groups}, 2019.
\newblock In press. \preprint{1611.00305}{math.QA}.

\bibitem{ZhaKL08}
L~Zhang.
\newblock Vertex tensor category structure on a category of
  {Kazhdan}-{Lusztig}.
\newblock {\em New York J. Math.}, 14:261--284, 2008.
\newblock \oldpreprint{0701260}{math.QA}.

\bibitem{HuaLeLiZha06}
Y-Z Huang, J~Lepowsky, H~Li, and L~Zhang.
\newblock On the concepts of intertwining operator and tensor product module in
  vertex operator algebra theory.
\newblock {\em J. Pure Appl. Algebra}, 204:507--535, 2006.
\newblock \oldpreprint{0409364}{math.QA}.

\bibitem{AstStr97}
A~Astashkevich.
\newblock On the structure of {Verma} modules over {Virasoro} and
  {Neveu}-{Schwarz} algebras.
\newblock {\em Comm. Math. Phys.}, 186:531--562, 1997.
\newblock \oldpreprint{9511032}{hep-th}.

\bibitem{RohRed96}
F~Rohsiepe.
\newblock On reducible but indecomposable representations of the {Virasoro}
  algebra.
\newblock \oldpreprint{9611160}{hep-th}.

\bibitem{RidSta09}
K~Kyt\"{o}l\"{a} and D~Ridout.
\newblock On staggered indecomposable {Virasoro} modules.
\newblock {\em J. Math. Phys.}, 50:123503, 2009.
\newblock \preprint{0905.0108}{math-ph}.

\bibitem{HuaYanLog12}
Y-Z Huang and J~Yang.
\newblock Logarithmic intertwining operators and associative algebras.
\newblock {\em J. Pure Appl. Algebra}, 216:1467--1492, 2012.
\newblock \preprint{1104.4679}{math.QA}.

\bibitem{DonVer98}
C~Dong, H~Li, and G~Mason.
\newblock Vertex operator algebras and associative algebras.
\newblock {\em J. Algebra}, 206:67--96, 1998.
\newblock \oldpreprint{9612010}{q-alg}.

\bibitem{HuaC209}
Y-Z Huang.
\newblock Cofiniteness conditions, projective covers and the logarithmic tensor
  product theory.
\newblock {\em J. Pure Appl. Algebra}, 213:458--475, 2009.
\newblock \preprint{0712.4109}{math.QA}.

\bibitem{HuaApp17}
Y-Z Huang.
\newblock On the applicability of logarithmic tensor category theory.
\newblock \preprint{1702.00133}{math.QA}.

\bibitem{CreBra18}
T~Creutzig, Y-Z Huang, and J~Yang.
\newblock Braided tensor categories of admissible modules for affine {Lie}
  algebras.
\newblock {\em Comm. Math. Phys.}, 362:827--854, 2018.
\newblock \preprint{1709.01865}{math.QA}.

\bibitem{PeaLog06}
P~Pearce, J~Rasmussen, and J-B Zuber.
\newblock Logarithmic minimal models.
\newblock {\em J. Stat. Mech.}, 0611:P11017, 2006.
\newblock \oldpreprint{0607232}{hep-th}.

\bibitem{ReaAss07}
N~Read and H~Saleur.
\newblock Associative-algebraic approach to logarithmic conformal field
  theories.
\newblock {\em Nucl. Phys.}, B777:316--351, 2007.
\newblock \oldpreprint{0701117}{hep-th}.

\bibitem{GaiLat13}
A~Gainutdinov and R~Vasseur.
\newblock Lattice fusion rules and logarithmic operator product expansions.
\newblock {\em Nucl. Phys.}, B868:223--270, 2013.
\newblock \preprint{1203.6289}{hep-th}.

\bibitem{GaiLog13}
A~Gainutdinov, J~Jacobsen, N~Read, H~Saleur, and R~Vasseur.
\newblock Logarithmic conformal field theory: A lattice approach.
\newblock {\em J. Phys.}, A46:494012, 2013.
\newblock \preprint{1303.2082}{hep-th}.

\end{thebibliography}
\providecommand{\oldpreprint}[2]{\textsf{arXiv:\mbox{#2}/#1}}
\providecommand{\preprint}[2]{\textsf{arXiv:#1 [\mbox{#2}]}}

\end{document}